\documentclass[11pt,a4paper]{article}
\usepackage{jcappub}
\usepackage{natbib}

\newcommand{\cR}{\,\mathrm{{\cal R}}}
\newcommand{\cN}{\,\mathrm{{\cal N}}}
\newcommand{\cG}{\,\mathrm{{\cal G}}}

\newcommand{\be}{\,\begin{equation}}
\newcommand{\ee}{\,\end{equation}}

\title{Diffusive propagation of cosmic rays from supernova remnants in the Galaxy. II: anisotropy}

\author{Pasquale Blasi and Elena Amato}

\affiliation{INAF/Osservatorio Astrofisico di Arcetri, Largo E. Fermi, 5 - 50125 Firenze, ITALY}

\emailAdd{blasi@arcetri.astro.it, amato@arcetri.astro.it}

\abstract{
In this paper we investigate the effects of stochasticity in the spatial and temporal distribution of supernova remnants on the anisotropy of cosmic rays observed at Earth. The calculations are carried out for different choices of the diffusion coefficient $D(E)$ experienced by cosmic rays during propagation in the Galaxy. The propagation and spallation of nuclei (with charge $1\leq Z\leq 26$) are taken into account.  At high energies ($E>1 $ TeV) we assume that $D(E)\propto (E/Z)^{\delta}$, with $\delta=1/3$ and $\delta=0.6$ being the reference scenarios. The large scale distribution of supernova remnants in the Galaxy is modeled following the distribution of pulsars with and without accounting for the spiral structure of the Galaxy. Our calculations allow us to determine the contribution to anisotropy resulting from both the large scale distribution of SNRs in the Galaxy and the random distribution of the nearest remnants. The naive expectation that the anisotropy amplitude scales as $\delta_{A}\propto D(E)$ is shown to be a wild oversimplification of reality which does not reflect in the predicted anisotropy for any realistic distribution of the sources. The fluctuations in the anisotropy pattern are dominated by nearby sources, so that predicting or explaining the observed anisotropy amplitude and phase becomes close to impossible. Nevertheless, the results of our calculations, when compared to the data, allow us to draw interesting conclusions in terms of the propagation scenario to be preferred both in terms of the energy dependence of the diffusion coefficient and of the size of the halo.
We find that the very weak energy dependence of the anisotropy amplitude below $10^{5}$ GeV, as observed by numerous experiments, as well as the rise at higher energies, can best be explained if the diffusion coefficient is $D(E)\propto E^{1/3}$. Faster diffusion, for instance with $\delta=0.6$, leads in general to an exceedingly large anisotropy amplitude. The spiral structure introduces interesting trends in the energy dependence of the anisotropy pattern, which qualitatively reflect the trend seen in the data. The inhomogeneous spatial distribution of the sources in the Galactic disc induces a large scale anisotropy which is not sensitive to the stochastic nature of nearby SNRs: we find that this additional contribution to $\delta_{A}$ becomes more important for large values of the size of the halo, $H$. The two terms are comparable in size for $H\sim 2$ kpc which corresponds to the scale height of the gradient of the spatial distribution of SNRs in the Galaxy. The dependence on energy of $\delta_{A}(E)$ is close to monotonic when the large-scale, regular term dominates, and does not seem to reflect the observed anisotropy amplitude. Both contributions to the total anisotropy are illustrated and discussed with the help of semi-analytical results.} 

\begin{document}
\maketitle

\section{Introduction}

The anisotropy of cosmic rays (CRs) may be a precious tool to probe the propagation of CRs throughout the Galaxy. A comprehensive understanding of the origin of CRs can only be achieved through a combined investigation of the spectrum, chemical composition and anisotropy. In paper I \cite{paper1} we have discussed the spectrum and chemical composition of CRs as observed at Earth when the sources are assumed to be supernova remnants (SNRs). The stochastic distribution of SNRs in the Galaxy was found to lead to interesting effects: if diffusion is parametrized in terms of a diffusion coefficient $D(E)\propto E^{\delta}$, the best fit to the all-particle spectrum is found to require $\gamma_{obs}=\gamma+\delta=2.67$, where $\gamma$ is the slope of the injection spectrum. In a rigidity dependent scenario for the CR acceleration and propagation, the knee in the all-particle spectrum is determined by the combination of the spectra of different chemicals. Qualitatively the resulting spectrum well reproduces the observed one, but the stochastic nature of the supernova events leads to fluctuations which imply that the naive expectation $\gamma_{obs}=\gamma+\delta$ does not hold strictly, a fact that makes it less straightforward to infer the actual injection spectrum required to fit the data. In an approximate sense, however, it remains true that a stronger energy dependence of the diffusion coefficient corresponds to harder injection spectra. This is a very important point in that the best known acceleration mechanism, diffusive shock acceleration (DSA) \cite{Bell:1978p1342,Bell:1978p1344,Blandford:1987p881}, invariably leads to spectra very close to power laws with slope $\gamma\simeq 2$ (naively leading to require $\delta\sim 0.6-0.7$). In fact, modern versions of DSA, including non-linear effects induced by the pressure of CRs in the shock region and by the super-Alfvenic streaming of CRs in the upstream region make two fundamental predictions: 1) concave spectra, possibly even harder than $E^{-2}$ at high energies, and 2) large values of the maximum achievable energy, plausibly as large as the knee for protons \cite{Blasi:2007p144}, and $Z$ times larger for nuclei with charge $Z$. The two aspects are strictly connected with each other: efficient acceleration leads to a high CR pressure, which in turn is responsible for a pronounced precursor and concave spectra; large CR acceleration efficiencies also imply magnetic field amplification and possibly higher maximum energies if CRs manage to interact resonantly with the self-generated magnetic perturbations. 

It has recently been proposed that the non-linear theory of DSA (hereafter NLDSA) may lead to somewhat steeper spectra if the velocity of the scattering centers in the amplified magnetic field is taken into account in solving the transport equation for CRs at SNR shocks \cite{Caprioli:2010p133,Caprioli:2010p789,Ptuskin:2010p1025}. In this case the spectra of accelerated particles may be less concave, and have an approximate power law shape with slope $\approx 2.1-2.2$. The steeper slope of the injection spectrum follows from both the finite velocity of the scattering waves and from the convolution over time of the acceleration history of SNRs throughout their evolution in the interstellar medium (ISM). However, there are two disappointing elements in this picture: first, these values of $\gamma$ would still imply $\delta \approx 0.5-0.6$, in order to ensure $\gamma_{obs}\sim 2.7$; in addition, the essential role played by the velocity of the scattering centers introduces much uncertainty since the results depend rather strongly on poorly understood details such as the wave helicity, as well as the reflection and transmission of waves at the shock surface (in principle the effect discussed here could even lead to harder spectra). Another disappointing ingredient, that has important consequences on the shape of the spectrum, is represented by the details of the magnetic field amplification, as recently discussed by \cite{Caprioli:2011p1915}. This again goes in the direction of steepening the spectra, although a non linear analysis of this effect has not been carried out as yet.

As shown in Paper I, one element that can help us discriminating the case of fast diffusion ($\delta\sim 0.6$, $\gamma\sim 2.1$) from slow diffusion ($\delta\sim 1/3$, $\gamma\sim 2.3$) is represented by the shape of the all-particle spectrum and the chemical composition: fast diffusion leads to sizeable fluctuations in the spectra, which woud imply that fitting the all particle spectrum depends upon the specific realization of the source distribution, invalidating any given physical criterion like $\gamma+\delta=\gamma_{obs}$. The case of fast diffusion also leads in general to negligible spallation of nuclei, so that the Helium spectrum at Earth is roughly parallel to that of protons, in contradiction with recent CREAM \cite{Ahn:2010p624} and PAMELA \cite{Adriani:2011p69} data that suggest a harder He spectrum above TeV energies. There might be different explanations for the hardening of the He spectrum (see e.g. \cite{ohiraioka:2011p729,vladimirov2011}), but it is however noticeable that if the diffusion coefficient is properly normalized and taken to scale with energy as $E^{1/3}$, this puzzling feature may be very naturally explained as a result of propagation from discrete sources. 

In Paper I we also discussed how the CR spectrum at Earth is affected by the details of the spatial and temporal distribution of SNRs. We showed that taking into account the spiral structure of the Galaxy leads to small changes compared to the case when the disk is taken as smooth. The changes become increasingly appreciable the smaller the assumed size of the magnetized galactic halo. We also considered different scenarios in terms of escape of the CRs from the acceleration sites and the possibility that supernovae of type II may preferentially be clustered within OB associations/superbubbles. We found that the resulting all-particle spectrum is very little sensitive to these assumptions. 

While the changes induced in the spectrum by the choice of $\delta$ or by the stochasticity in the spatial and temporal distribution of the sources are small, they become very large when the anisotropy amplitude is considered, a fact that makes this quantity very interesting to investigate. This is the purpose of the present paper. 

The two benchmark cases, $\delta=1/3$ and $\delta=0.6$ lead to very different predictions for the anisotropy, as already noticed in Ref. \cite{Ptuskin:2006p620}: the anisotropy obtained in the case of fast diffusion ($\delta=0.6$) is in general much larger than the observed one. Moreover the fluctuations related to the specific realization of sources are very large, making it impossible to obtain a quantitative prediction to compare with data. The comparison must necessarily be qualitative. 

The fluctuations appear, though with different strengths, in all scenarios that we discuss in the following, and they play a crucial role in explaining the observed anisotropy. The naive prediction of the diffusion model with a continuous distribution of sources is that the anisotropy should increase with energy exactly as $E^{\delta}$. This is the type of anisotropy signal that can be calculated in numerical approaches such as GALPROP or DRAGON, which however can only single out the anisotropy signal due to the large scale inhomogeneous distribution of the sources (and/or to the off-center position of the solar system in the Galaxy). In fact, the anisotropy pattern of individual realizations shows a rather irregular behavior with local dips and peaks in some energy ranges, or sometimes showing approximate constancy over large energy intervals, similar to what the data show. The theoretically predicted scaling of the anisotropy with the diffusion coefficient can only be recovered when the size of the halo is taken to be larger than 4 kpc, in agreement with the fact that when the halo is large enough the large scale inhomogeneity of the source distribution becomes increasingly dominant over the small scale fluctuations due to nearby sources.

The paper is organized as follows: in \S~\ref{sec:green} we summarize the Green function formalism introduced in Paper I; in \S~\ref{sec:simple2} we derive simple estimates of the effect of fluctuations on the spectrum for a homogeneous but discrete distribution of sources in the disc. Our results on anisotropy for different models of source distribution and escape mechanisms of CRs from the sources are presented in \S~\ref{sec:results}. In \S~\ref{sec:Hdep} we discuss the dependence of the expected anisotropy signal on the size of the halo. We conclude in \S~\ref{sec:conclusion}.

\section{Summary of the Green functions formalism}\label{sec:green}

In this section we briefly summarize the Green function formalism that has been introduced in Paper I.
The diffusive transport of cosmic rays from a point source located at a position $\vec r_{s} = (x_{s},y_{s},z_{s})$ and injecting a spectrum $N(E)$ at a time $t_{s}$ can be written as:
\be
\frac{\partial n_{k}(E,\vec r,t)}{\partial t}=\nabla\left[D_{k}(E)\nabla n_{k}(E,\vec r,t)\right] - \Gamma_{k}^{sp}(E) n_{k}(E,\vec r,t) + N_{k}(E) \delta(t-t_{s})\delta^{3}(\vec r - \vec r_{s}),
\label{eq:transport}
\ee
where $n_{k}(E,\vec r,t)$ is the density of particles of type $k$ (nuclei) with energy $E$ at the location $\vec r$ and time $t$, $D_{k}(E)$ is the diffusion coefficient assumed to be spatially constant and $\Gamma_{k}^{sp}(E) $ is the rate of spallation of nuclei of type $k$ to lead to lighter nuclei. Each source is assumed to produce nuclei of H ($k=1$), He ($k=2$), CNO ($k=3$), Mg-Al-Si ($k=4$) and Fe ($k=5$). All injection spectra are assumed to be in the form:
\be
N_{k}(E) \propto E^{-\gamma} \exp \left[ -\left( \frac{E}{E_{max,k}} \right)\right],
\ee
where the normalization and the maximum energy of each species are chosen as discussed in Paper I. The maximum energies scale with the charge $Z$ of the nucleus.

The diffusion coefficient is taken in the form:
\be
D(E) = 10^{28} D_{28} \left( \frac{R}{3 GV}\right)^{\delta} \rm cm^{2} s^{-1} 
\label{eq:diff}
\ee
for rigidity $R>3$ GV, the secondary to primary ratios and the abundances of unstable isotopes can be best fit by choosing $D_{28}/H_{kpc}= 1.33$ for $\delta=1/3$ and  $D_{28}/H_{kpc}= 0.55$ for $\delta=0.6$, where $H_{kpc}$ is the height of the halo in units of kpc.

The diffusion region is assumed to be a cylinder of infinite radius and half height $H$, inside which the diffusion coefficient is spatially constant. The fact that the radius is taken potentially infinite is not a limitation to the calculation because our source distributions are concentrated in a region of a few kpc radius (the Galactic disc) and for practical purposes the size of the diffusion region can be taken to be much larger than the radius of the disc. The escape of CRs from this model Galaxy occurs through the upper and lower boundaries. The escape is modeled by assuming that $n_{k}$ vanishes at $z=\pm H$, so that the escape flux through the surfaces $z=\pm H$ is described by $D_{k}(E) \frac{\partial n_{k}}{\partial z}|_{z=\pm H}$. 

In this paper, as in Paper I, we concentrate our attention on particles with energy above $\sim 1$ TeV, therefore in Eq.~\ref{eq:transport} we ignored terms of advection and terms of possible second order reacceleration which are both potentially important at much lower energies. We also neglect, in computing the flux of nuclei, the contribution of secondaries produced in spallation events. From the point of view of how diffusive propagation is treated, the calculation described here is not much different from GALPROP or similar propagation codes. The only two relevant differences are in the more phenomenological way in which spallation is described and in the fact that here we do not impose the free escape boundary condition at some finite radius along the lateral sides of the cylinder. The latter, as discussed above, is not a relevant limitation, although it might lead to some corrections for large values of $H$. As to spallation, its rate is defined in terms of the process cross section and of the density of gas in the diffusion region as explained in Paper I. This treatment is adequate as long as only primary nuclei are considered, as we do here, while a more refined treatment would be required if we were to describe rare nuclei, such as $B$ and $^{10}Be$ or similar secondary products. 

The Green function that satisfies the correct boundary condition at $z=\pm H$ can be obtained through the image charge method and can be written as follows:
$$
{\cal G}_{k}(\vec r, t; \vec r_{s},t_{s}) = \frac{N_{k}(E,t_s)}{\left[ 4 \pi D_{k} \tau \right]^{3/2}} \exp\left[ -\Gamma_{k}^{sp}(E) \tau \right]
\exp\left[ -\frac{(x-x_{s})^{2}+(y-y_{s})^{2}}{4 D_{k} \tau}\right] \times
$$
\be
\sum_{n=-\infty}^{+\infty} (-1)^{n} \exp \left[ -\frac{(z-z'_{n})^{2}}{4 D_{k} \tau}\right],
\label{eq:green}
\ee
where $z'_{n}=(-1)^{n} z_{s} + 2 n H$ are the $z$ coordinates of the image sources. It is easy to check that ${\cal G}_{k}(x,y,z=\pm H,t;x_{s},y_{s},z_{s},t_{s})=0$ by simply expanding the sum term. In what follows, the Earth will be located at $(x,y,z)\equiv (R_{\odot},0,0)$, where $R_{\odot}=8.5$ kpc is the distance of the Sun from the center of the Galaxy. It is however useful to keep Eq.~\ref{eq:green} in its most general form because anisotropies are related to the spatial derivatives of the Green function calculated at the detection location. 

It is important to realize that the Green function formalism outlined here allows us to take into account a completely arbitrary spatial distribution of the sources in the Galaxy. The same is true for the temporal evolution of the injection of CRs by each source: one can introduce an arbitrary time dependence of the particle spectrum injected per unit time, $Q(E,t)$, and simply integrate Eq.~\ref{eq:green} over time.

We already discussed the importance of this latter point in Paper I, where we worked out the expression of $N(E,t)$ for the two scenarios that are most relevant for CR acceleration by SNRs. These are: 1) "burst injection", with all particles from a SNR injected at the same time with a spectrum $N(E)$; 2) continuous injection, with $Q(E,t)\propto \delta(E-E_{max}(t))$, namely with particles at the maximum energy escaping the SNR at all times after the beginning of the Sedov-Taylor phase, at time $T_{ST}$. In this latter scenario, the maximum energy the SNR can provide is $E_M$ and is reached at the time $T_{ST}$. Later on, the maximum energy decreases with time allowing particles of progressively lower energy to leave the system. The whole process lasts $\sim (3-30)\times 10^{4}$ years, and the convolution over time of these peaked spectra leads to a power law injection spectrum \cite{Caprioli:2009p145} which is not directly related to the spectrum of particles at the shock. The global injected spectrum from an individual SNR is likely to be the result of the superposition of the spectrum of particles escaping the SNR from upstream and of those that escape after the end of the Sedov-Taylor phase. The latter component, however, is heavily affected by adiabatic energy losses, and therefore it does not contribute to the energy region around the knee.

In the first scenario, namely for "burst injection", the Green function in Eq.~\ref{eq:green} is already the solution that we are seeking, with
\be
N_{k}(E) = \frac{(\gamma_{k}-2) \eta_{k}\epsilon_{kin}}{E_{0,k}^{2}} \frac{1}{\left[ 1-\left( \frac{E_{max,k}}{E_{0,k}}\right)^{-\gamma_{k}+2}\right]} \left(\frac{E}{E_{0}} \right)^{-\gamma_{k}} \exp\left( - \frac{E}{E_{max,k}}\right),
\label{eq:inj}
\ee
where $\eta_{k}$ is the fraction of the kinetic energy of the blast wave, $\epsilon_{kin}$, that goes into accelerated particles of type $k$. The reference energy $E_{0,k}$ is assumed to be $1$ GeV for protons (for heavier nuclei its numerical value is not really important since we simply rescale the injected protons spectrum in order to fit the spectra observed at Earth). In Eq.~\ref{eq:inj}, $E_{max,k}$ is the maximum energy of particles of type $k$. In the expression above and in what follows we always assume that the injection spectrum is steeper than $E^{-2}$, since flatter spectra would result in unreasonable choices for the energy dependence of the diffusion coefficient (leading to exceedingly large anisotropy and even to breaking the regime of diffusive propagation, see below). 

For the second model of injection, we assume that the maximum energy of the accelerated particles increases during the ejecta dominated phase of the remnant evolution to reach a maximum at the beginning of the Sedov-Taylor phase, and start decreasing afterwards. Particles at the instantaneous maximum energy escape the SNR at all times $t\geq T_{ST}$ \cite{Caprioli:2009p145}, carrying a fraction $\eta(t)$ of the explosion energy. We take the decrease with time of $E_{max}$ in the form of a power-law, $E_{max}(t)=E_M(t/T_{ST})^{-\alpha}$, with $\alpha>0$ chosen in such a way as to guarantee that $E_{max}(\tau_{SNR})=E_0$, where $\tau_{SNR}$ is the time at which the SNR dies out and we take $E_0=1 GeV$ for protons. With these assumptions, and using the Sedov-Taylor description of the shock dynamics we find for the injected spectrum:
\be
Q(E,t)=\eta(t) \frac{\epsilon_{kin}}{E_M} \frac{1}{T_{ST}} \left(\frac{6}{5}\right)^3 \left(\frac{t}{T_{ST}}\right)^{\alpha-1}~~~~~~~t>T_{ST}\ .
\label{eq:qe}
\ee
We further assume $\eta(t)=\eta_0(t/T_{ST})^\beta$, with $\beta>0$, as discussed in Paper I. Supplied with this latter condition, the expression in Eq.~\ref{eq:qe} can be used in Eq.~\ref{eq:green}. Integration over time then leads to (see Paper I for details): 

$$
n(E,\vec r,t) = \eta_{0} ~\epsilon_{kin} \left( \frac{6}{5}\right)^{3} \frac{1}{\alpha E_{M}^{2}}
\left( \frac{E_{M}}{E} \right)^{2} \left( \frac{E_{M}}{E} \right)^{\frac{\beta}{\alpha}} \times
$$
$$
\frac{1}{\left[ 4 \pi D(E) \tau^{*} \right]^{3/2}} \exp\left[ -\Gamma_{k}^{sp}(E) \tau^{*} \right]
\exp\left[ -\frac{(x-x_{s})^{2}+(y-y_{s})^{2}}{4 D_{k} \tau^{*}}\right] \times
$$
\be
\sum_{n=-\infty}^{+\infty} (-1)^{n} \exp \left[ -\frac{(z-z'_{n})^{2}}{4 D_{k} \tau^{*}}\right],
\label{eq:sol1}
\ee
where 
$$
\tau^{*}=t-t_{s}-T_{ST}\left( \frac{E_{M}}{E} \right)^{\frac{1}{\alpha}}.
$$
From Eq.~\ref{eq:sol1} it is clear that the solution is non-zero only for 
\be
1\leq \left( \frac{E_{M}}{E} \right)^{\frac{1}{\alpha}} \leq \rm Min\left[ 1+\frac{\tau_{SNR}}{T_{ST}},\frac{t-t_{s}}{T_{ST}}\right].
\label{eq:ineq}
\ee
Of all the inequalities in Eq.~\ref{eq:ineq}, the interesting case is that of recent SNRs, for which $(t-t_s)/T_{ST}< 1+\tau_{SNR}/T_{ST}$. In this case, at time $t$, the SNR only contributes particles with energies
\be
E> E_M \left(\frac{t-t_s}{T_{ST}}\right)^{-\alpha}\ .
\label{eq:lowen}
\ee
This lower limit on energy comes from the fact that lower energy particles are still confined in the source. This feature is peculiar to the scenario of continuous injection and is completely unrelated to propagation effects, while the latter set, in all cases, an additional, independent bound on the minimum energy of CRs reaching us from a given source.

\section{Anisotropy and its fluctuations}\label{sec:simple2}

The main goal of the present paper is to discuss the implications of diffusive CR propagation from discrete sources on the anisotropy signal measured at Earth. In a regime of diffusive propagation, the anisotropy in a given direction is defined as 
\be
\delta_{\vec x} = \frac{3 D(E)}{c} \frac{\nabla_{\vec x} n_{CR}(E,\vec r,t)}{n_{CR}},
\label{eq:anisdef}
\ee
where $n_{CR}(E)$ is the CR number density as measured at the position of the Earth. The density gradient, as calculated over many realizations of sources, has two contributions, one that is sensitive to the mean distribution of sources over large spatial scales, and one that is sensitive to fluctuations induced by random nearby sources. Whether one or the other dominates the anisotropy signal depends in general upon the spatial scales on which the gradients in the source distribution appear. 

It is easy to see that for a homogeneous distribution of sources in the disc, the mean value of $\delta_{\vec x}$ vanishes in all directions, simply because there are no gradients in the disc ($z=0$) in such a situation, except for boundary effects close to the edge of the disc. Even in this situation, however, the measured anisotropy will be non-zero, due to the small scale inhomogeneity of the source distribution in the solar neighborhood. It is clear that this contribution will not show the scaling with  $D(E)$ that one expects based on Eq.~\ref{eq:anisdef}. For an inhomogeneous distribution of sources, as the one that we think appropriate to describe SNRs in the Galaxy, it is not straightforward to predict what the anisotropy will be as a function of energy, since this depends on whether the small or large scale contribution will dominate. In order to quantify these effects we follow here the argument of \cite{Lee:1979p1621}, already used in Paper I to calculate the fluctuations in the spectral shape of CRs at Earth as due to the stochastic distribution of sources.

In order to write the anisotropy it is useful to introduce the particle current. In the case of diffusive propagation, this reads:
\be
\vec J_{CR}(\vec r,t) = -D \vec \nabla n_{CR}(\vec r,t)\ .
\label{eq:partcurr}
\ee
The overall anisotropy will then result from the sum of two terms, representing the large-scale and small-scale contribution respectively. These can be written in terms of $\vec J_{CR}$ as:
\be
\delta_{A1}=\frac{\langle \vec J_{CR}\rangle}{\frac{1}{3}cn_{CR}(E)}
\label{eq:delt1}
\ee
and
\be
\delta_{A2}=\frac{\langle \delta \vec J_{CR}  \cdot \delta \vec J_{CR}\rangle^{1/2}}{\frac{1}{3}c n_{CR}(E)}\ .
\label{eq:delt2}
\ee

If we consider now $\cN$ independent sources, each producing $N_0$ particles, following the formalism of \cite{Lee:1979p1621} and of Paper I, we can write the average particle current entering Eq.~\ref{eq:delt1} as:
\be
\langle\vec J_{CR}(\vec r,t)\rangle =N_0 \cN D(E)\int d^{3}\vec r' dt' 
\vec \nabla_r \cG (\vec r,t;\vec r',t') P(\vec r',t')\ ,
\label{eq:avgJ}
\ee
while for its fluctuations, entering Eq.~\ref{eq:delt2}, we have:
\be
\langle \vec \delta J_{CR}(\vec r,t) \vec \delta J_{CR}(\vec r,t)\rangle={N_0}^2 \cN D(E)^2 \int_d^{3}\vec r'' dt'' 
\vec \nabla_r \cG (\vec r,t;\vec r'',t'') \cdot \vec \nabla_{r'} \cG (\vec r',t';\vec r'',t'') 
P(\vec r'',t'')\ .
\label{eq:avgdJ}
\ee

In the previous expressions $P(\vec r, t)$ is the probability of having a source at the position $(\vec r,t)$ at time $t$, $\cG(\vec r,t;\vec r',t')$ is the Green function for transport of particles from $(\vec r',t')$ to $(\vec r,t)$ and terms of ${\cal O}(1/\cN)$ have been neglected.

Now let us consider a homogeneous spatial distribution of the sources in the disc, described by the probability function
\be
P(\vec r, t)=\frac{{\cal R}}{\pi R_d^2}\ .
\label{eq:punif}
\ee
In this case, it is easy to show that: 
$\langle \vec J_{CR}(\vec r,t) \rangle=0$ and hence the large scale term in Eq.~\ref{eq:delt1} is zero.
As to the small scale term, for purpose of illustration, let us compute it for the case of protons alone, so that spallation can be ignored, and let us consider an infinitely thin Galactic disc, so that $\langle \vec \delta J_{CR} \cdot \vec \delta J_{CR}\rangle=\langle \delta J_{CR,x} \delta J_{CR,x}\rangle + \langle \delta J_{CR,y} \delta J_{CR,y}\rangle$. This case is clearly much simpler than the scenarios we actually considered in our calculations, but it still allows us to discuss a few key features.
From Eq.~\ref{eq:avgdJ} we have:
$$\langle \delta J_{CR} \delta J_{CR} \rangle = \frac{\cR}{\pi {R_d}^2} \int_{T_{min}}^\infty \frac{d\tau}{(4\pi d\tau)^3} \int_0^{R_d} 
\frac{r^2}{(2D\tau)^2} \exp \left[-\frac{r^2}{2D\tau}\right] \times$$
\be
\times\sum_{m=-\infty}^\infty \sum_{l=-\infty}^\infty (-1)^{l+m} \exp\left[-\frac{(l^2+m^2)H^2}{D\tau}\right]\ ,
\label{eq:fluct1}
\ee
where both integrals above can be calculated analytically, and, if we assume $H^{2}/D(E)T_{min}\gg 1$, the only important term in the sums is easily seen to be that corresponding to $l+m=0$. The result for the rms current is then:
\be
\langle \delta J_{CR} \delta J_{CR}\rangle \simeq \frac{1}{128\ \pi^3} \frac{\cR N_{0}^{2}}{D(E) T_{min}^{2} R_d^2}.
\label{eq:deltadelta}
\ee
At the same time, it is straightforward to calculate the average CR density (see Paper I). For $H\ll R_d$, this can be written as:
\be
\langle n_{CR}\rangle=\frac {N(E)\cR}{2\pi R_d^2} \frac{H}{D(E)}\ .
\label{eq:dens}
\ee

Using the expressions in Eq.~\ref{eq:deltadelta} and Eq.~\ref{eq:dens}, from Eq.~\ref{eq:anisdef} we finally obtain:
\be
\delta_{A2} = \frac{3}{2^{5/2}\pi^{1/2}}\frac{D(E)^{1/2}}{c \cR^{1/2}}\left( \frac{R_{d}}{H}\right) \frac{1}{T_{min}}.
\label{eq:anis}
\ee

First thing to notice in Eq.~\ref{eq:anis} is the strong dependence of the small scale anisotropy (in terms of normalization and in terms of energy dependence) on the cutoff time $T_{min}$, which we were forced to impose in order to avoid the divergence in the integral over $\tau$. One can see that, if $T_{min}$ is naively chosen as a given number, the anisotropy is a very slow function of energy, $\delta_{A2}\sim E^{\delta/2}$. On the other hand, the time $T_{min}$ should carry information about the closest, most recent sources around the observer. For instance the time $T_{min}$ could be interpreted as the time over which one source goes off within a distance from Earth such that CRs from that source reach us within a time $T_{min}$. This condition is expressed as
\be
\frac{\cR T_{min}}{\pi R_{d}^{2}} \left[4 \pi D(E)T_{min}\right] = 1 \to T_{min} = \left[\frac{4 \cR D(E)}{R_{d}^{2}}\right]^{-1/2}.
\ee
With this prescription the small-scale anisotropy reads \cite{Ptuskin:2006p620}:
\be
\delta_{A2} = \frac{3}{2^{3/2}}\frac{1}{\pi^{1/2}} \frac{D(E)}{H c}, 
\label{eq:finanis}
\ee
which shows an energy dependence $\delta_{A2}\sim E^{\delta}$. Moreover $\delta_{A2}\propto D(E)/H$, namely a scaling that is the inverse of that of the CR flux, $n_{CR}(E)\propto H/D(E)$. The fact that both quantities depend upon the ratio of $D$ and $H$, and that the same ratio at, say, 10 GeV, is fixed by measurements of the B/C ratio, implies that the assumed size of the halo has little effect in changing the small-scale anisotropy or the spectrum. However this statement depends on the specific choice of $T_{min}$ that we have made and cannot be taken as general: when fluctuations are important, the actual anisotropy can easily be expected to depend strongly upon the specific distribution of nearby and most recent sources.

On the other hand, for any realistic distribution of sources in space, the zero-th moment of the anisotropy, $\delta_{A1}$, does not vanish and corresponds to a large scale anisotropy due to the simple fact that on average there may be more sources in one direction (towards the Galactic center) than in the opposite direction. The first moment, $\delta_{A2}$, as calculated above refers to the fluctuations on top of this non-vanishing mean anisotropy and whether it dominates or not again depends on the specific distribution of sources. For the source density profiles that will be used in the sections below, though the energy dependence of the amplitude of anisotropy is approximately the same for both terms, the scaling with the halo size $H$ is different. The anisotropy arising from the overall inhomogeneous source distribution has a faster dependence on $H$. For small values of $H$, comparable with the large scale gradient in the source distribution, $\delta_{A1}$ is comparable in magnitude to $\delta_{A2}$, but becomes increasingly dominant with increasing $H$. Although the energy dependence of the mean anisotropy over many source realizations for the two terms is similar for our assumed $T_{min}$ (both proportional to $D(E)$), when $\delta_{A2}$ is negligible compared to $\delta_{A1}$ the anisotropy amplitude approaches a monotonic function of energy. At the same time, the phase of the anisotropy is very different: it correlates strongly with the direction of the Galactic center for large $H$ while it is a stochastic variable for small $H$, a symptom of the dominance of nearby recent SNRs in this latter case. The importance of nearby sources in terms of anisotropy was also discussed in \cite{2006APh....25..183E} where the authors adopt a stochastic generation method for SNRs in the Galaxy similar to the one used in this paper.

\section{CR anisotropy for realistic distributions of SNRs}\label{sec:results}

To first approximation, the anisotropy of CRs that is observed at Earth results from the combination of two main physical ingredients: the global spatial distribution of SNRs in the Galaxy, which in turn leads to gradients in the CR distribution, and the proximity effect of nearby and/or recent supernova events. In general other elements can also affect the anisotropy, like, for instance, longer residence times of CRs close to their sources resulting from the self-generation of turbulence. 

In this section we summarize the two models of spatial distribution of SNRs, already discussed in Paper I, the {\it cylindrical model} and the {\it spiral model}. The cylindrical model is basically that discussed, for instance, in \cite{Case:1996p1635}. The other model (hereafter "spiral model") is instead an attempt at taking into account the spiral distribution of sources in the Galaxy: for this we adopt the formalism of \cite{FaucherGiguere:2006p1609}.

In both cases the sources are assumed to have a mean radial distribution in the Galaxy following the function \cite{Case:1996p1635}:
\be
f(r)=\frac{A}{R_{\odot}^{2}}\left( \frac{r}{R_{\odot}}\right)^{2} \exp \left[ -\beta\frac{r-R_{\odot}}{R_{\odot}}\right],
\label{eq:radial}
\ee
where $\beta=3.53$ for $R_{\odot}=8.5$ kpc. 

The constant $A$ is determined from the normalization condition:
\be
\int_{0}^{\infty} dr~2\pi r f(r) = 1 \to A=\frac{\beta^{4}\exp(-\beta)}{12 \pi}.
\ee
The SNR distribution in the $z$ direction is assumed to be 
\be
f(z) = \frac{A_{z}}{z_{g}} \exp\left[ -\frac{z}{z_{g}}\right],
\ee
where $A_{z}=1$ is again derived from the normalization condition. 

In the cylindrical model, the positions of the sources are chosen by drawing at random values of $r$ and $z$ from the distributions above. The $x$ and $y$ coordinates are chosen by generating a random angle $0\leq \phi \leq 2\pi$ and using the given value of $r$. Supernovae are generated at a rate $\cR=1/(30-100)~ \rm yr^{-1}$, and the generation of new sources is continued until a time span much larger than $H^{2}/D(E)$ is covered, in order to make sure that the stationary solution has been reached at the lowest rigidities of interest for us ($\sim 1$ TV). Typically in our calculations we choose a time span of $80$ million years, more than sufficient to ensure stationarity at rigidities larger than $1$ TV. 

In the {\it spiral model}, the procedure we adopt is similar to that introduced in \cite{FaucherGiguere:2006p1609} and summarized in Paper I: the generation of the position in the $z$ direction is the same as above, therefore we will not discuss it any further. A radial coordinate $\tilde r$ is drawn at random from the average distribution in Eq.~\ref{eq:radial}; at this point a random natural number between 1 and 4 is chosen from a flat distribution. This number identifies the arm in which the supernova is localized (Norma, Carina-Sagittarius, Perseus, Crux-Scutum, as in Table 1 of Paper I). At this point an angular position along the arm is associated to the SNR, according to the relation:
\be
\theta(r) = K \log \left(\frac{r}{r_{0}}\right) + \theta_{0}.
\ee
The parameters $K$, $r_{0}$ and $\theta_{0}$ are reported in Table 1 of Paper I.

Following the prescription of \cite{FaucherGiguere:2006p1609}, we blur the angle $\theta(r)$ by $\theta_{corr}\exp(-\tilde r[{\rm kpc}]/w[{\rm kpc}])$: here $\theta_{corr}$ is chosen from a flat random distribution between 0 and $2\pi$, while for $w$ we consider two reference values, $w=2.8$ and $w=5$ kpc and study how the results are affected by this choice. The distribution proposed in Ref. \cite{FaucherGiguere:2006p1609} has $w=2.8$ kpc, but the choice of the value of this parameter does not appear to be strongly data motivated. 

Similarly to the angular position, the radial coordinate is also blurred by choosing a {\it new} value of the radius from a normal random distribution with mean $\tilde r$ and variance $0.07\tilde r$ \cite{FaucherGiguere:2006p1609}. 

\begin{figure}[t]
\centering\leavevmode
\includegraphics[width=3.in,angle=0]{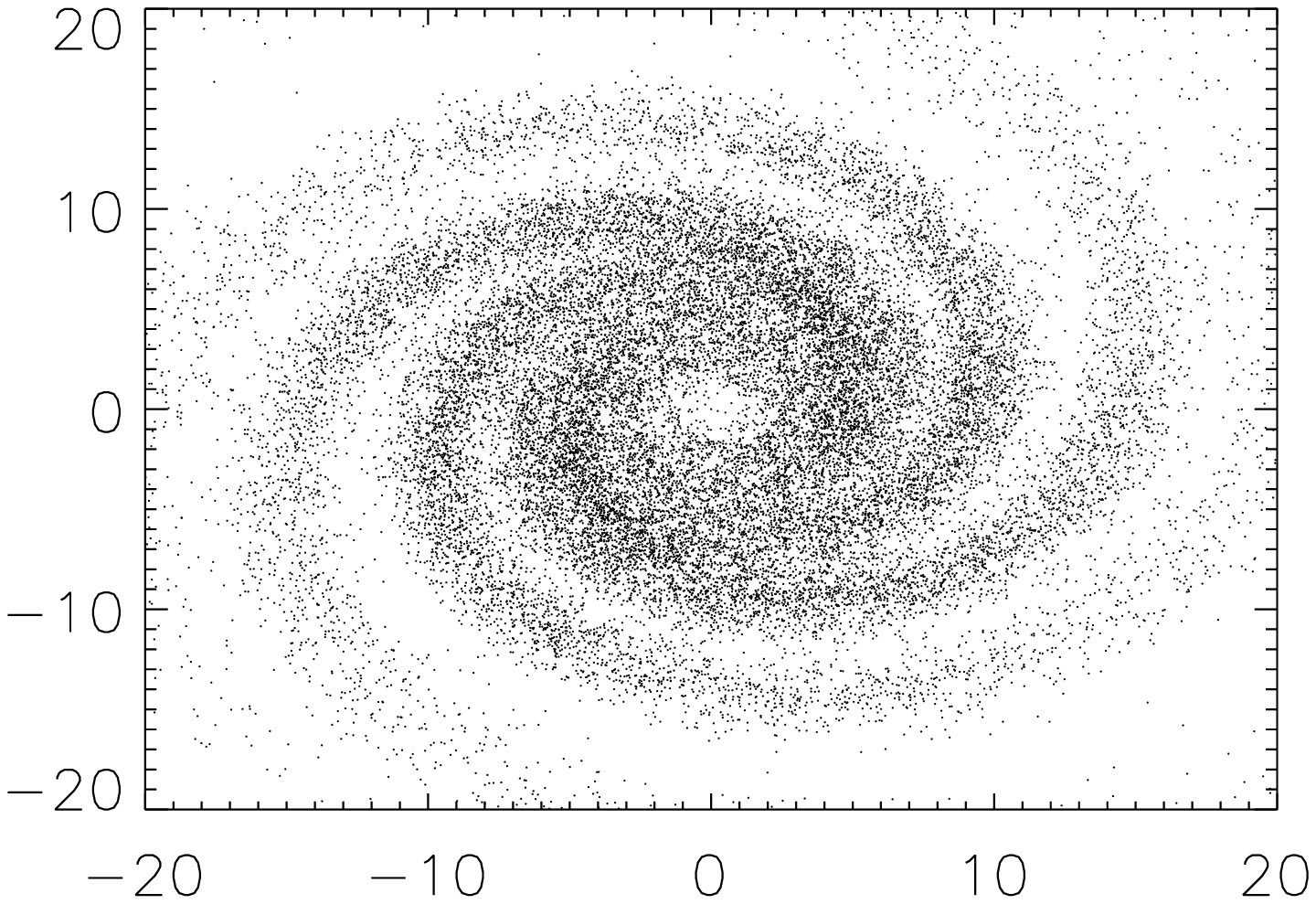}
\includegraphics[width=3.in,angle=0]{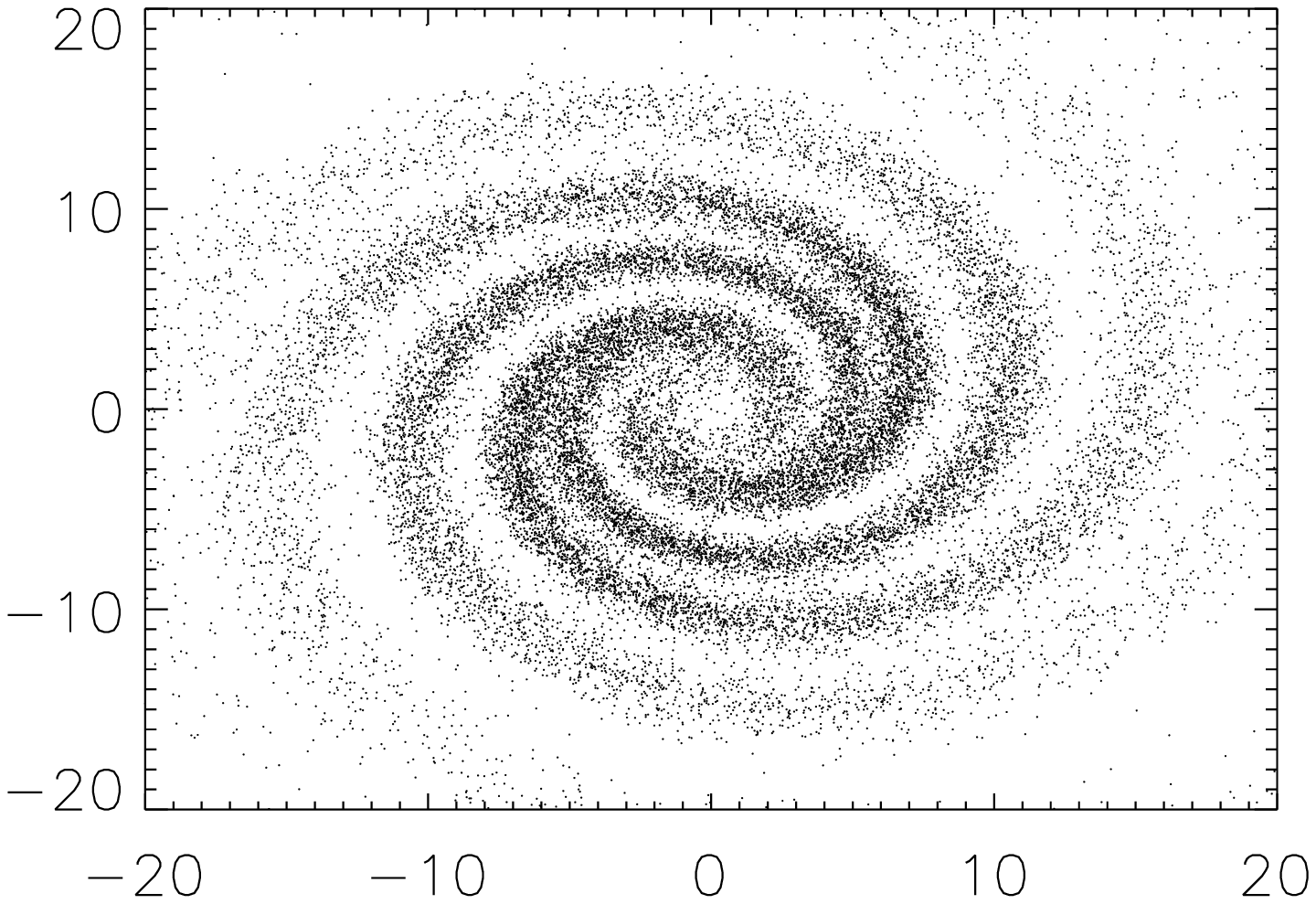}
\caption{A face-on view of the spatial distribution of SNRs in the Galaxy for two different models of the spiral structure: $w=5$ kpc on the left and $w=2.8$ kpc on the right. About $3\times 10^4$ sources are shown in each panel. Units are in $kpc$ and the position of the Sun is at $(x,y)=(R_{\odot},0)$.}  
\label{fig:space}
\end{figure}

In Fig.~\ref{fig:space} we show how the spiral structure changes depending on the value of $w$: the figure shows the distribution of $\sim 30,000$ SNRs in a case where $w=5$ kpc (left) and in a case with $w=2.8$ kpc (corresponding to the value adopted by \cite{FaucherGiguere:2006p1609}). The position of the Sun is at $(x,y)=(R_{\odot},0)$.

For each realization of source distributions in the two models the contribution to anisotropy is calculated in all directions, so that both a magnitude and a phase can be evaluated. 

\subsection{Anisotropy for the cylindrical model}\label{sec:res_anis}

\begin{figure}[t]
\centering\leavevmode
\includegraphics[width=2.8in,angle=0]{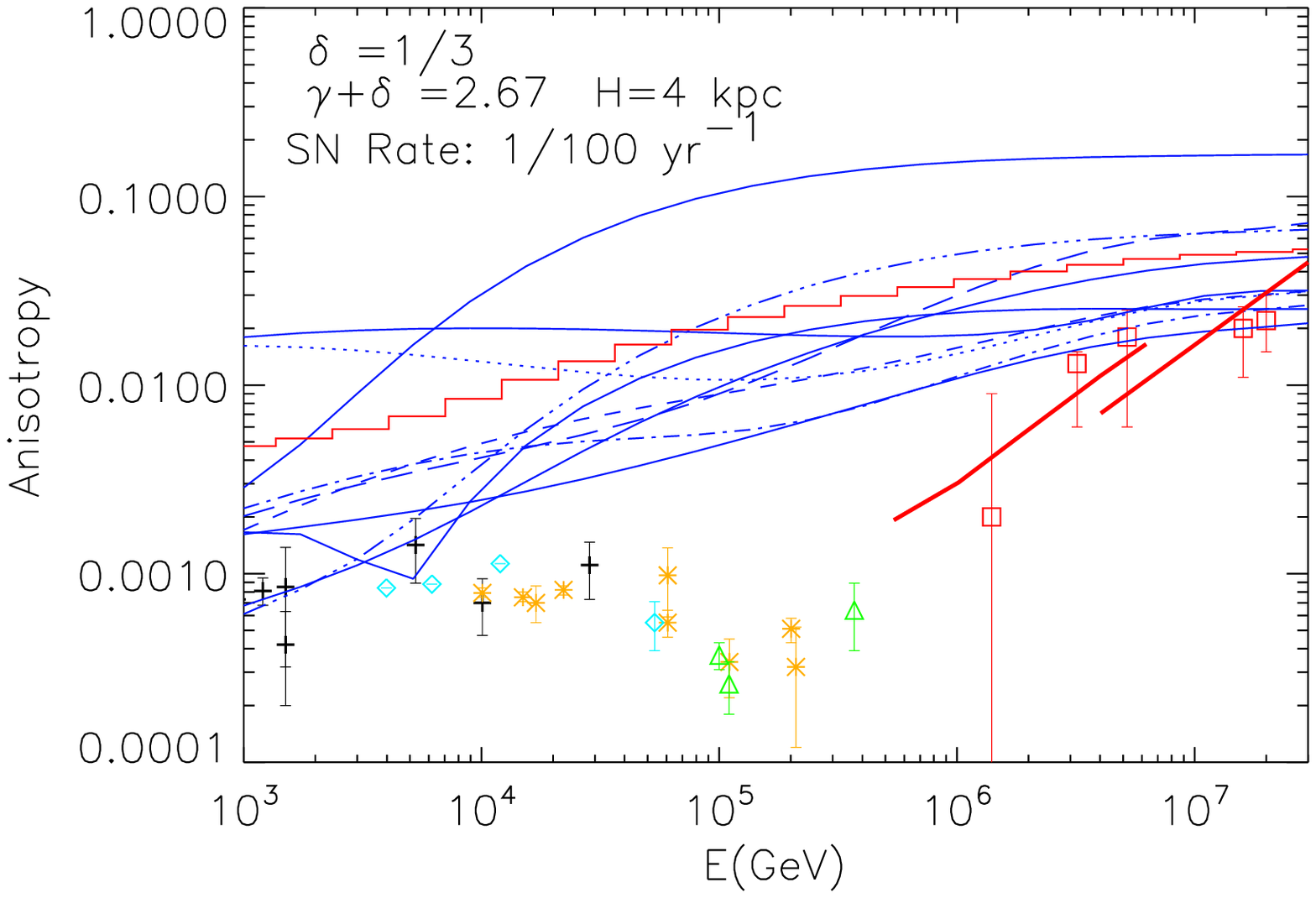}
\includegraphics[width=2.8in,angle=0]{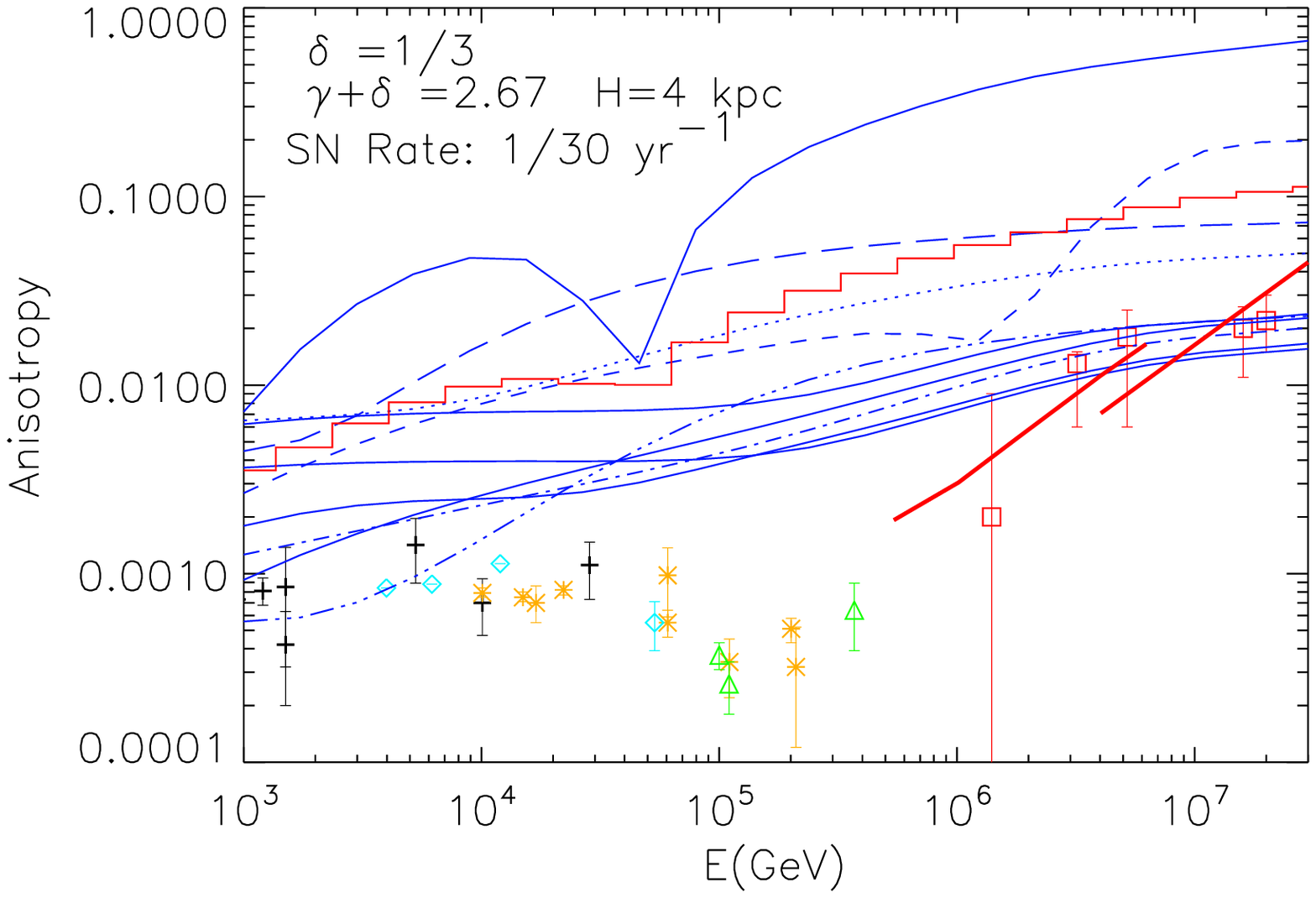}
\caption{Anisotropy amplitude for ten random realizations of sources in the cylindrical model, assuming $\delta=1/3$ and a SN rate $\cR=1/100$ yr$^{-1}$ ($\cR=1/30$ yr$^{-1}$) on the left (right). The halo size is $H=4$ kpc. The injection spectrum is assumed to have slope (below the cutoff) such that $\gamma+\delta=2.67$. The data points are from \cite{Amenomori:2005p1770,Aglietta:2003p1913,1986JPhG...12..129K}.}  
\label{fig:varyR}
\end{figure}

In this section we discuss the anisotropy signal for the case of cylindrical symmetry in the SNR distribution. 
In Fig.~\ref{fig:varyR} we illustrate the anisotropy amplitude for 10 realizations of the source distribution in the cylindrical model, using $\delta=1/3$, $H=4\ kpc$ and a rate of supernova explosions in the Galaxy $\cR=1/100$ yr$^{-1}$ (left) and $\cR=1/30$ yr$^{-1}$ (right). In all cases we impose that the slope $\gamma$ of the injection spectrum is related to $\delta$ through $\gamma+\delta=2.67$, in order to ensure a good fit to the CR spectrum at Earth (see Paper I). The red, staircase line represents the average amplitude calculated using the 10 random realizations. 

In all figures the (black) crosses, the (blue) diamonds and the (orange) stars are taken from Ref.~\cite{Amenomori:2005p1770}. The (green) triangles are from EASTOP \cite{Aglietta:2003p1913,Aglietta:2009p130} and the (red) squares are the Akeno data points \cite{1986JPhG...12..129K}. The oblique (red) lines at high energy show the upper limits on the amplitude of anisotropy from KASCADE and GRANDE \cite{Antoni:2004p665}.

The comparison between the two panels shows that the spread in the anisotropy patterns is not affected in a significant way by the SN rate. This can be qualitatively understood if one considers that for $H=4$ kpc, the anisotropy signal is already dominated by $\delta_{A1}$ (see \S~\ref{sec:Hdep}). Looking at Eq.~\ref{eq:avgJ} one sees that the rate of Supernova explosions $\cR$ only enters $\langle J_{CR}\rangle$ (and the same is true for $n_{CR}$) through the normalization of the probability distribution. It is then clear that any dependence on $\cR$ will disappear when $\delta_{A1}$ is obtained as the ratio between $\langle J_{CR}\rangle$ and $n_{CR}$.
Both panels of Fig.~\ref{fig:varyR} show very clearly the strong dependence of the strength of anisotropy on the specific realization of source distribution, thereby also disproving the naive expectation that the anisotropy should be a growing function of energy with the same slope as the diffusion coefficient $D(E)$. Whenever the small scale contribution is not negligible, the observed anisotropy can in fact even be a non monotonic function of energy, with dips and bumps, and with wide energy regions in which it is flat with energy, quite like what the data show at energies $E<10^{5}$ GeV. It is interesting however that none of our realizations of the source distribution leads to anisotropies as low as the one suggested by the data in the energy region $10^5-10^6$ GeV (contributed by the EASTOP experiment). 

Data in this region are in fact somewhat puzzling because they are so low as to suggest that the Compton-Getting effect \cite{gleesonaxford68} leads to a level of anisotropy close to the lowest expected limit. The Compton-Getting anisotropy is estimated to be between $3\times 10^{-4}$ and $10^{-3}$ depending on the velocity with which the Earth moves with respect to the rest-frame of the CR scattering centers. This velocity is not known and the above estimates refer to a velocity range from a minimum of $\sim 20$ km/s to a maximum of $\sim 250$ km/s, corresponding to the motion of the solar system through the Galaxy \cite{ambrosioetal03}. It is clear that the measured anisotropy between $10^5$ and $10^6$ GeV is only marginally consistent with a velocity of few tens of km/s at most.

We also checked the effects of decreasing further the source rate, which could be the case if the bulk of CRs does not come from standard SNe but rather from rarer events, like for example an especially energetic sub-sample of SNe or GRBs. The resulting anisotropy is somewhat larger at low energies, on average: the data can still be easily reproduced at the low and high energies, but the central, more problematic region is now more extended, in general, to the left than in Fig.~\ref{fig:varyR}, approximately ranging from few $\times 10^4$ to $10^6$. 

In Fig.~\ref{fig:varyR} we adopted a diffusion coefficient scaling with $E^{1/3}$. The energy dependence of the diffusion coefficient is however the subject of an ongoing debate: given $D(E)\propto E^\delta$ it is controversial whether $\delta$ is $1/3$, $1/2$, 0.6 or even larger (see \cite{maurinetal2010} and references therein).

\begin{figure}[t]
\centering\leavevmode
\includegraphics[width=4.8in,angle=0]{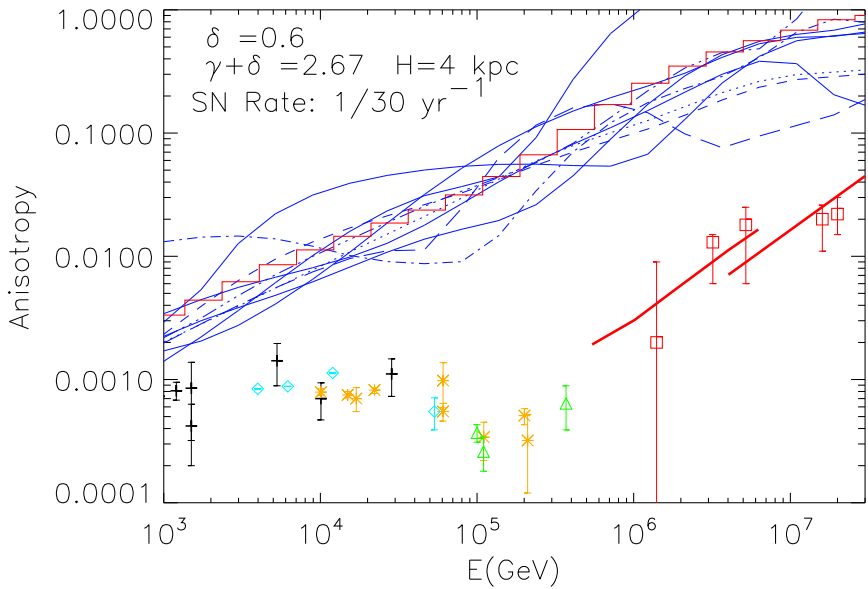}
\caption{Anisotropy amplitude for ten random realizations of sources in the cylindrical model, assuming $\delta=0.6$ and a SN rate $\cR=1/30$ yr$^{-1}$. The halo size is $H=4$ kpc. The injection spectrum is assumed to have slope (below the cutoff) such that $\gamma+\delta=2.67$.}  
\label{fig:varydel}
\end{figure}

The all-particle spectrum alone, while giving some indications that $\delta=1/3$ could be preferable (see Paper I), does not allow one to really clinch the question. This is because the all-particle spectrum only depends on the combination $\delta+\gamma$. In principle the B/C ratio would allow a direct measurement of $\delta$, if this ratio could be measured at sufficiently high energies. Unfortunately at the present time the error bars on this quantity are still large enough to allow for ambiguity in the best fit value (see for instance \cite{Ahn:2008p1594}).

Since the anisotropy $\delta_A$ is defined as the ratio between the density gradient and the density, $\gamma$ does not appear in $\delta_A$ while $\delta$ does (see also expressions \ref{eq:anis} and \ref{eq:finanis} for the simplified case of a uniform distribution of the sources). In Fig.~\ref{fig:varydel} we plot the amplitude of the anisotropy computed for ten different realizations of the source distribution in the cylindrical model: a slope of the diffusion coefficient $\delta=0.6$ is assumed, while the other parameters are all the same as for the plot in the right panel of Fig.~\ref{fig:varyR}. 

As well as in the case $\delta=1/3$, also for $\delta=0.6$ the amplitude of the anisotropy is a complex function of energy as a result of the cosmic rays contributed by nearby recent SNRs. However, for $\delta=0.6$ the amplitude of the anisotropy appears to be systematically larger than the observed one at all energies. In other words, fast diffusion leads to exceedingly large anisotropy which seriously challenges the models that require large values of $\delta$ (see for instance the discussion in Ref.~\cite{maurinetal2010} for the cases in which a convective wind is included). It is worth noticing that at very high energies the amplitude may exceed unity. These cases clearly suggest that the diffusive paradigm may break down for very nearby sources of CRs, as already discussed in Paper I. 

We think that the results just showed provide clear evidence in favor of a diffusion coefficient with a weak dependence on energy. This finding is of crucial importance in several respects. The fact that the data suggest a value $\delta=1/3$ is comforting in some respects and puzzling in some others, in relation to our understanding of CR acceleration and propagation. On the one hand, $\delta=1/3$ gives the exact energy dependence of $D(E)$ that Kolmogorov-type turbulence would provide, so propagation follows a framework that was not unpredicted from the theoretical point of view. On the other hand, however, as we already mentioned in Paper I, $\delta=1/3$ implies that the injection spectrum should be a power law with slope $\gamma\sim2.3-2.4$ which is much steeper than what acceleration theory has been predicting for the last several years, especially due to the non linear effects which make the spectrum concave. We defer further discussion of this point to \S~\ref{sec:conclusion}, while below we discuss the dependence of anisotropy on other unknowns of the problem, like the real source distribution, the size of the halo and the detailed time-dependence of the injection of CRs by each source. 

The results discussed so far are obtained within the assumption that each source releases accelerated particles in a time that is short compared to all the relevant time-scales of the problem; moreover, the released particles are assumed to have a power-law spectrum. As already anticipated it is wise to also consider a different scenario, involving time-dependent injection (see \S~\ref{sec:green}). In this model, the release of CRs in the ISM occurs continuously during the Sedov-Taylor phase of expansion of the SNR, with higher energy particles being released earlier (close to the beginning of the Sedov phase) and particles of gradually decreasing energy at correspondingly later times. The adopted dependence on time and energy of the injected particle distribution is that illustrated in \S~\ref{sec:green}, Eq.~\ref{eq:qe} and Eq.~\ref{eq:sol1}. It is however necessary to keep in mind that this recipe should not be considered as a realistic attempt to model the complex escape of CRs from a SNRs, an objective that is way beyond the purpose of the present paper and that has been discussed at length elsewhere (see {\it e.g.} \cite{Caprioli:2009p145}). Our purpose here is only to explore the effect of gradual injection of CRs into the ISM on the amplitude of anisotropy.

\begin{figure}[t]
\centering\leavevmode
\includegraphics[width=2.8in,angle=0]{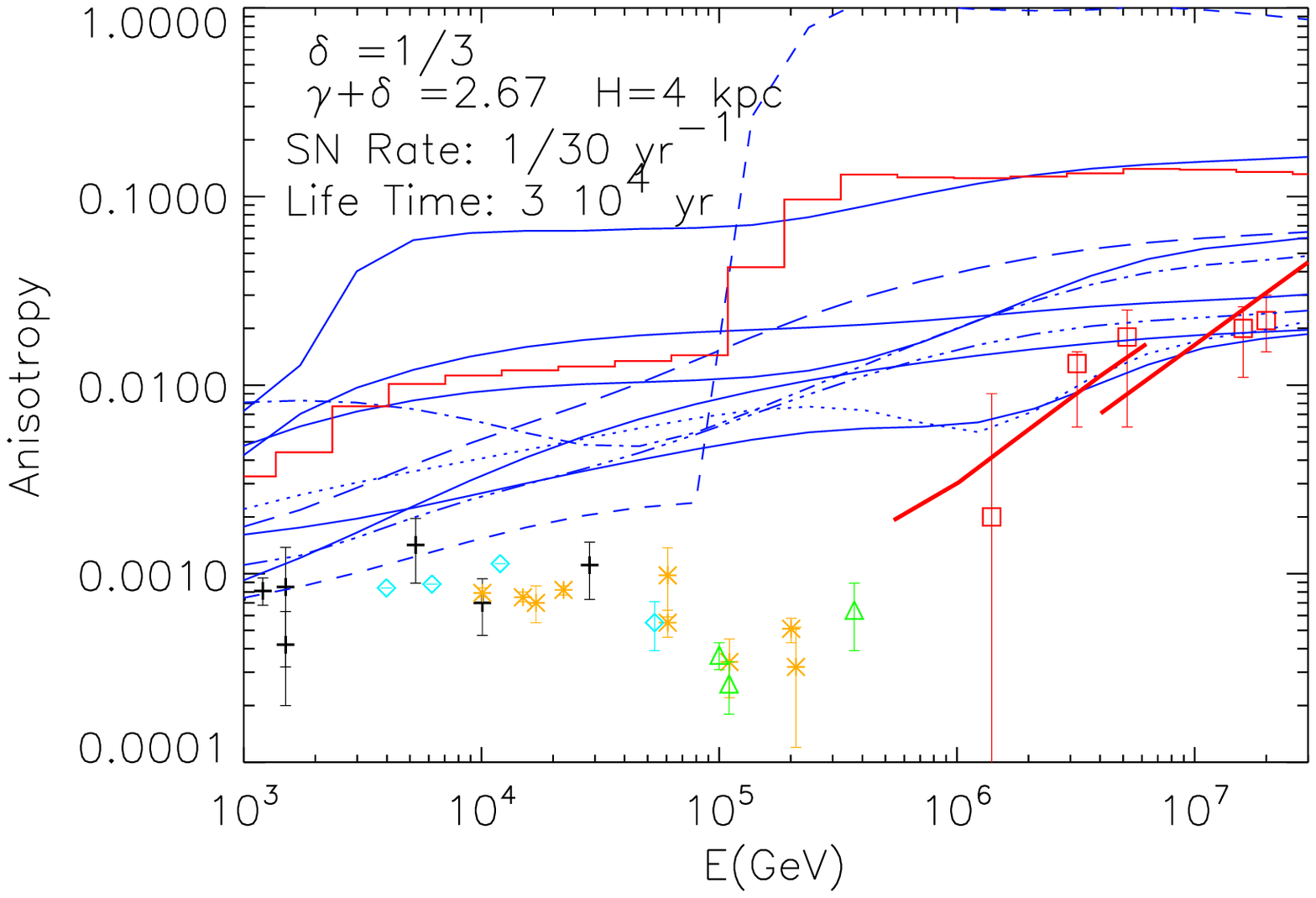}
\includegraphics[width=2.8in,angle=0]{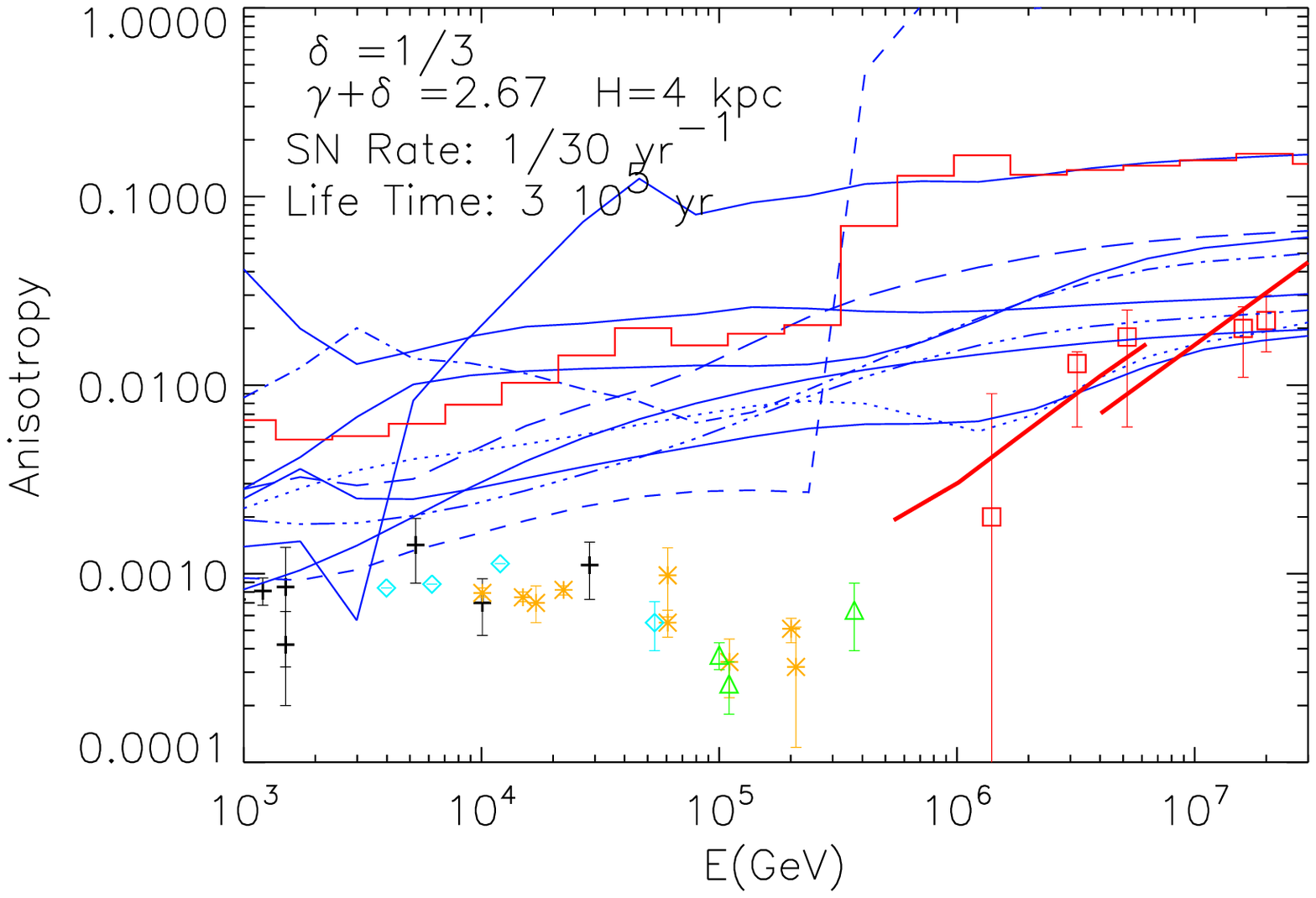}
\caption{Anisotropy amplitude for ten random realizations of sources in the cylindrical model in the case of continuous escape of particles from the upstream of the accelerating blast wave. The two different panels are relative to different assumptions on the SNR lifetime: $\tau_{SNR}=3 \times 10^4$ yr on the left;  $\tau_{SNR}=3 \times 10^5$ yr on the right. Other parameters are: $\delta=1/3$, a SN rate $\cR=1/30$ yr$^{-1}$ and a halo size $H=4$ kpc. The injection spectrum is such as to ensure $\gamma+\delta=2.67$, see text for details.}
\label{fig:varylife}
\end{figure}

In Fig.~\ref{fig:varylife} we plot the anisotropy resulting from ten different configurations for two different choices of the SNR lifetime, $\tau_{SNR}=3 \times 10^4$ on the left and $\tau_{SNR}=3 \times 10^5$ on the right. The resulting curves are qualitatively similar to those obtained for bursting injection of CRs into the ISM, with possibly some additional bumpiness caused by the fact that some nearby sources did not have time yet to generate the low energy particles. It remains true that there are several realizations of source distributions which appear to provide a reasonable fit to the observed anisotropy amplitude. The data points between $10^5$ and $10^6$ GeV remain somewhat peculiar.

The two different durations of the Sedov-Taylor phase do not make much difference, once the parameters are adjusted in such a way as to guarantee that the final slope of the observed spectrum, $\gamma_{obs}$, is the same: this requires that once $\alpha$ is determined from the condition $\alpha=\log(E_{M}/E_0)/\log(\tau_{SNR}/T_{ST})$, $\beta$ is given as $\beta=\alpha(\gamma_{obs}-2-\delta)$. 

\subsection{CR anisotropy in the spiral model}

In this section we discuss how the anisotropy of CRs is affected by the spiral structure of the Galaxy. 

\begin{figure}[t]
\centering\leavevmode
\includegraphics[width=2.8in,angle=0]{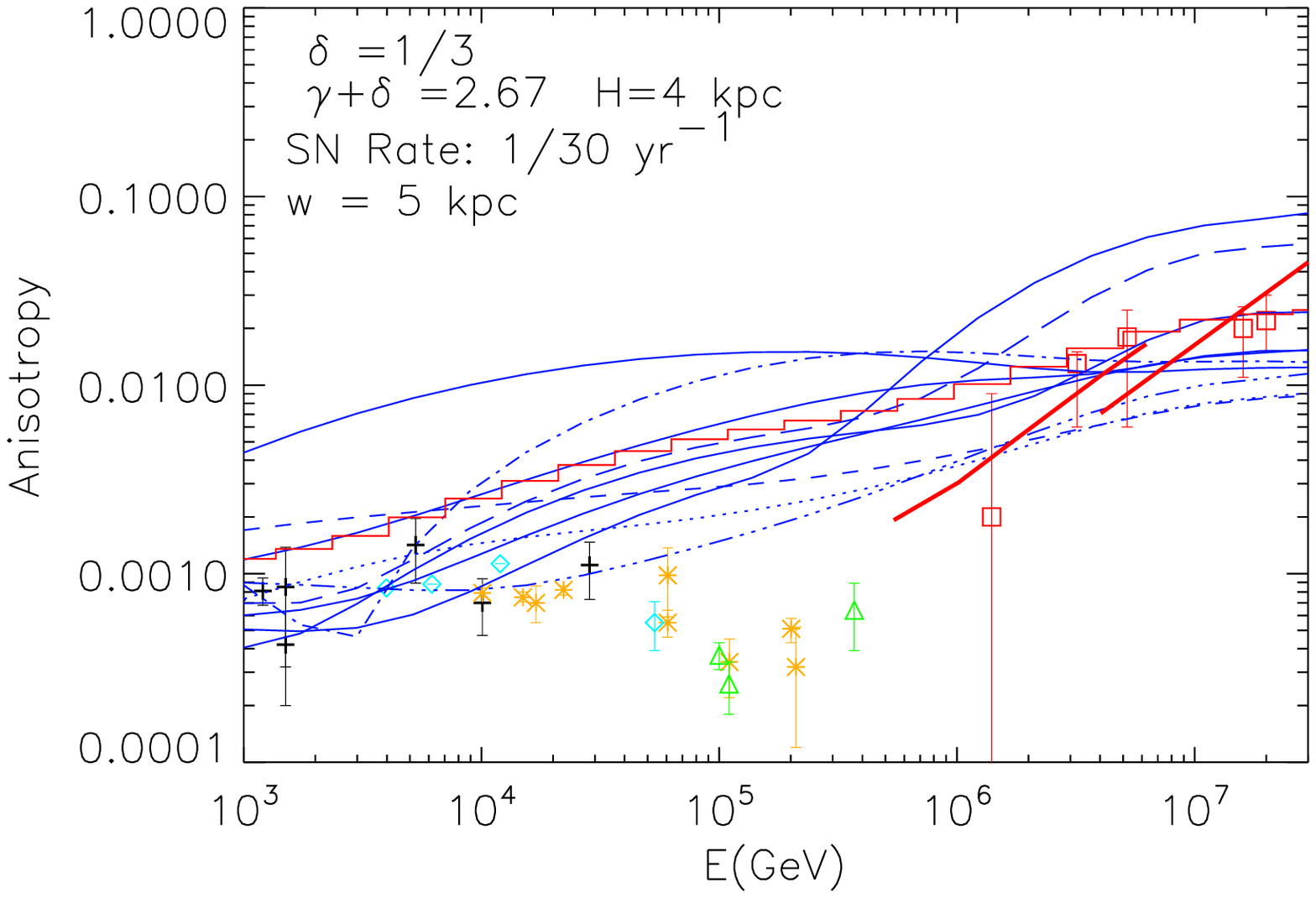}
\includegraphics[width=2.8in,angle=0]{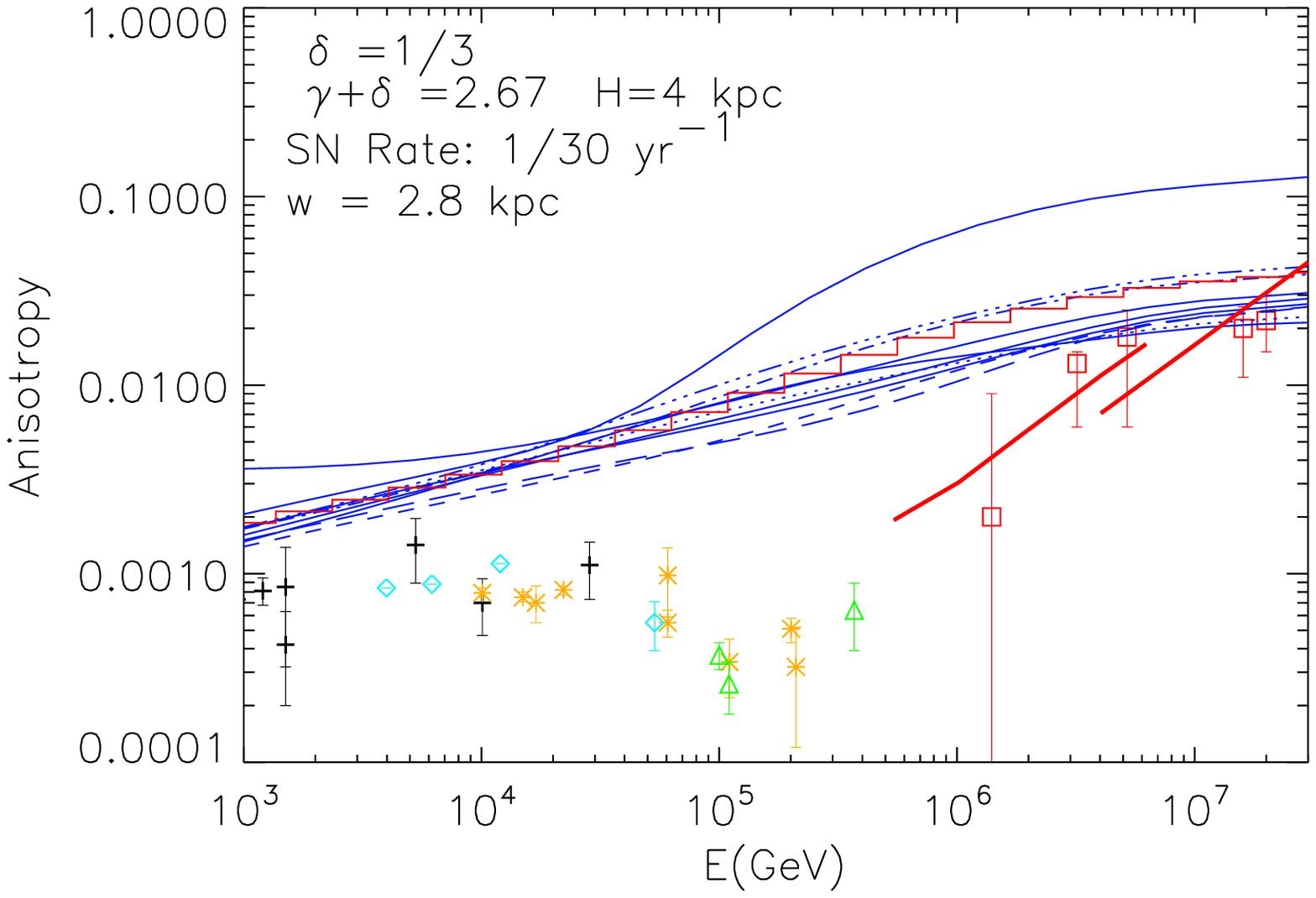}
\caption{Anisotropy amplitude for ten random realizations of sources in the spiral model. The curves in the two panels are obtained for different values of $w$, with $w=5$ kpc on the left and $w=2.8$ kpc on the right (source distributions as in the left and right panel of Fig.~\ref{fig:space} respectively). Bursting injection is considered. Other parameters are: $\delta=1/3$, a SN rate $\cR=1/30$ yr$^{-1}$ and a halo size $H=4$ kpc. The injection spectrum is such as to ensure $\gamma+\delta=2.67$, see text for details.}
\label{fig:varyw}
\end{figure}

In Fig.~\ref{fig:varyw} we plot the anisotropy computed for a number of different realizations of the sources' distribution in the spiral model. The two panels relate to two different choices for the parameter $w$ introduced above, the same two values for which the spatial distribution of the sources is plotted in Fig.~\ref{fig:space}, $w=2.8$ kpc and $5$ kpc. 
First thing we notice in Fig.~\ref{fig:varyw} is that in the left panel ($w=5$ kpc) the dispersion among the different curves is much larger than on the right ($w=2.8$ kpc). The reason for the different shape of the curves for the anisotropy amplitude as a function of energy for the two values of $w$ is related to the large scale structure of the spiral: as we discussed several times, the CR density gradient is determined by both the large scale distribution of SNRs and by the accidental proximity of nearby recent sources. Whether the anisotropy is dominated by one or the other effect depends upon whether the halo size $H$ is larger or smaller than the spatial scale on which gradients in the source distribution appear. In the case of a spiral with a tight spread ($w=2.8$ kpc), the large scale distribution of the sources is such that the nearby recent SNRs are subdominant and in fact the curves showing the amplitude of anisotropy are very regular and monotonically increasing with energy. In the case $w=5$ kpc, the sources are more spread out around the centroid of the arms and it becomes more likely to have recent SNRs located in the proximity of the Sun's location, therefore the anisotropy shows the bumps and dips that we have already discussed in the previous section. 

\begin{figure}[t]
\centering\leavevmode
\includegraphics[width=2.8in,angle=0]{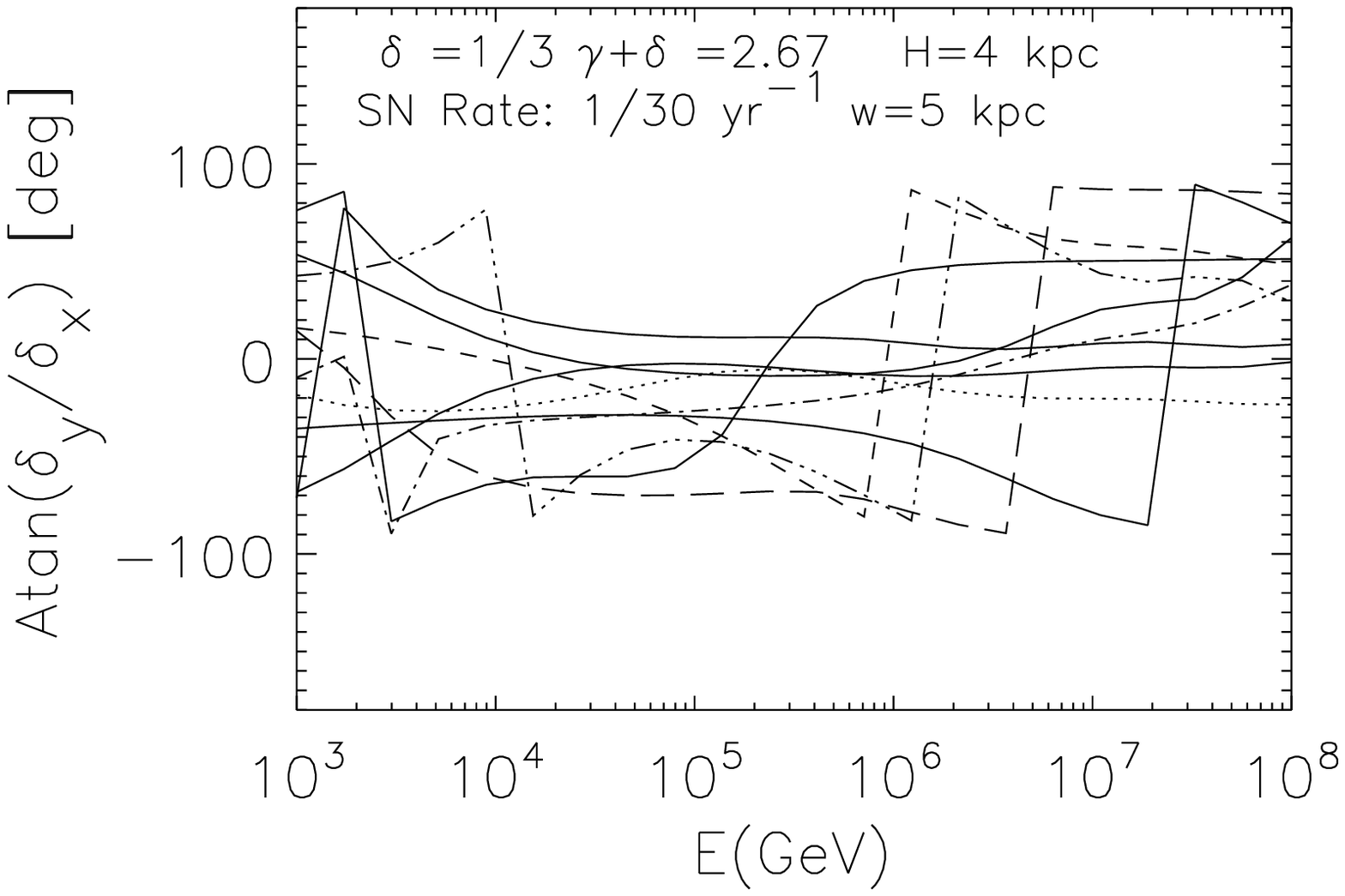}
\includegraphics[width=2.8in,angle=0]{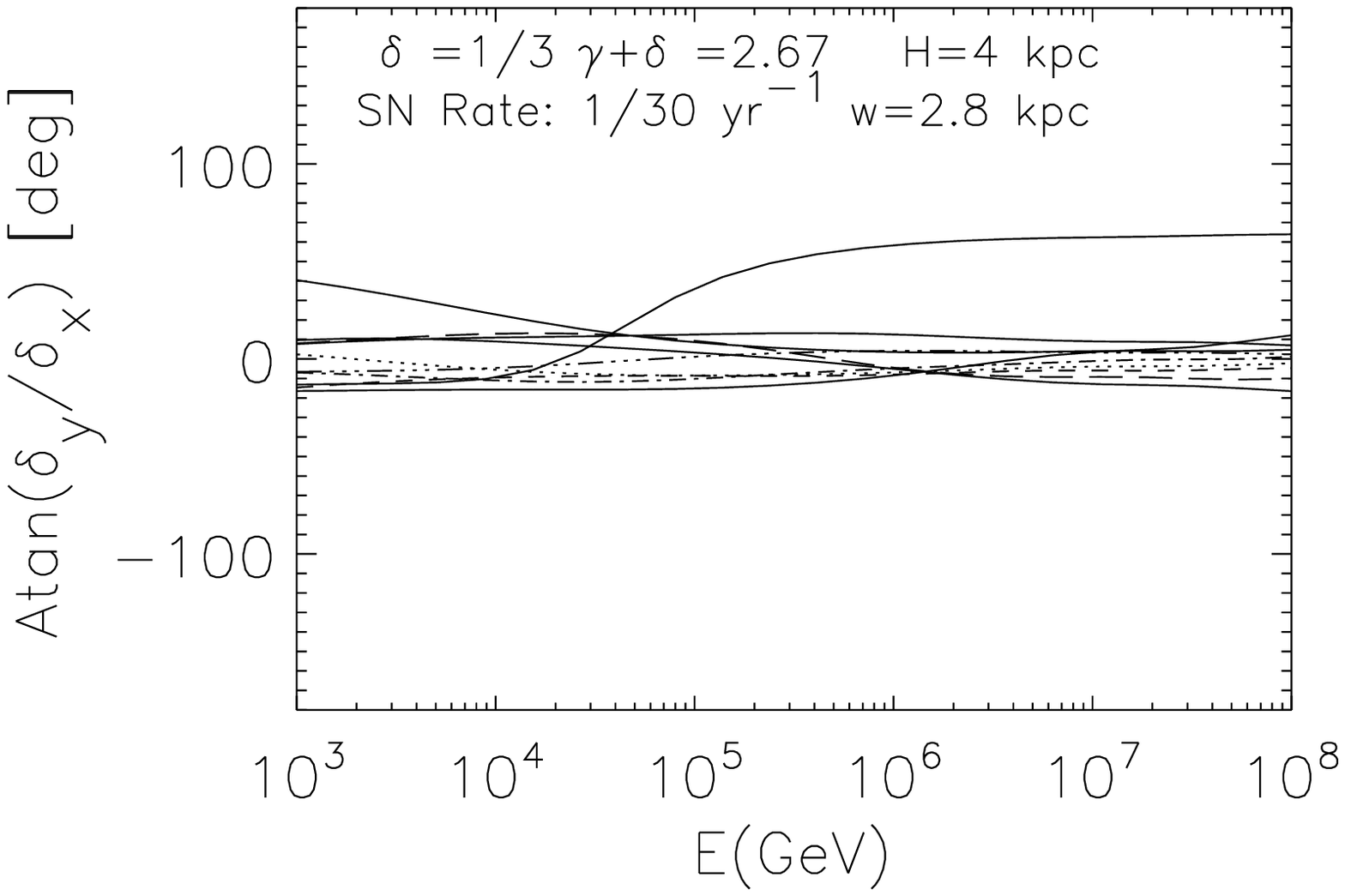}
\caption{Phase of the anisotropy for the same cases as in the left and right panels of Fig. \ref{fig:varyw}.}
\label{fig:phase}
\end{figure}

This interpretation of the results of our calculations is also confirmed by the investigation of the phase of the anisotropy vector. In Fig.~\ref{fig:phase} we plot $\arctan(\delta_{y}/\delta_{x})$ (phase of the anisotropy vector in the $xy$ plane of the Galactic disc) as a function of energy. The left (right) panel refers to $w=5$ kpc ($w=2.8$ kpc). As expected, in the case of narrow arms ($w=2.8$ kpc) the anisotropy is dominated by the large scale distribution of sources and the phase has little spread around zero. On the other hand, for broader arms ($w=5$ kpc) the fluctuations induced by recent nearby SNRs dominate the anisotropy amplitude and the phase has a wide spread depending on the realization of the source distribution. 

It is worth stressing that the phase of the anisotropy as derived by our calculations shows instances of abrupt changes in contiguous energy bins, as a result of the dominance, in those bins, of two (or more) different sources (the same behaviour is found in the case of cylindrical distribution of sources). This type of behaviour is reminiscent of that found by \cite{Aglietta:2009p130} in the two energy bins $E=1.1\times 10^{14}$ eV and $3.7\times 10^{14}$ eV. In the same paper the authors also tried to make sense of the slope of the amplitude of the anisotropy between the same energy bins, and found a value $\delta\sim 0.74$, which they interpreted in terms of energy dependence of the CR diffusion coefficient. As showed by our calculations, such a slope is unlikely to have anything to do with the diffusion coefficient, and is rather the result of the large fluctuations induced by local sources. 

\section{The dependence of the anisotropy amplitude on the size of the halo $H$}
\label{sec:Hdep}

As we discussed in \S~\ref{sec:simple2}, for a homogeneous distribution of sources in a thin disc the only contribution to the anisotropy is the one due to the proximity of local recent sources. In fact, the non central position of the Sun in the disc also gives rise to a small anisotropy but this component is negligible compared to that associated with fluctuations. Moreover, the anisotropy as derived in \S~\ref{sec:simple2} (see Eq. \ref{eq:finanis}) is independent of the size $H$ of the halo, once the constraint $D_{28}/H_{kpc}\sim 1$ is taken into account. It follows that making the size of the halo bigger should not lead to any appreciable change in the anisotropy amplitude. 

This situation changes however when the realistic distribution of sources in the disc is taken into account. 

\begin{figure}
\centering\leavevmode
\includegraphics[width=4.8in,angle=0]{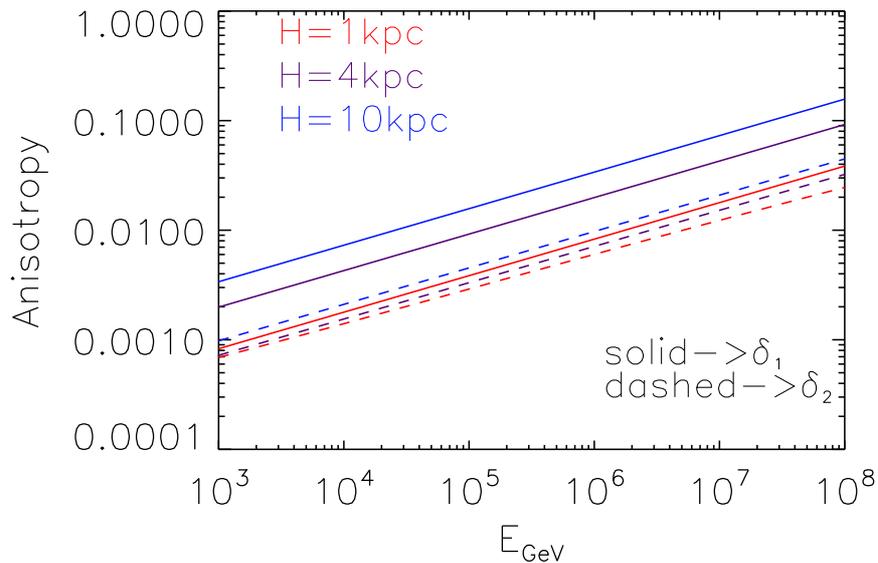}
\caption{Relative importance of the large scale ($\delta_{A1}$, solid lines) and small-scale ($\delta_{A2}$, dashed lines) contribution to the anisotropy signal. The two terms are plotted for different sizes of the halo ($H=1$ kpc (red), $H=2$ kpc (purple), $H=10$ kpc (blue). The diffusion coefficient is assumed to scale as $E^{1/3}$.}
\label{fig:analytic}
\end{figure}

We computed the terms $\delta_{A1}$ and $\delta_{A2}$ according to Eqs.~\ref{eq:delt1} and \ref{eq:delt2} where in the definition of $\langle \vec J_{CR} \rangle$ and $\langle \delta J_{CR} \delta J_{CR}\rangle$ (Eqs.~\ref{eq:avgJ} and \ref{eq:avgdJ}) we used the probability distribution described by Eq.~\ref{eq:radial} and normalized to the supernova rate: in practice we computed the anisotropy corresponding to a continuous (instead of stochastic) version of the cylindrical model for source distribution. The results of this calculation are plotted in Fig.~\ref{fig:analytic}.

It is clear from this figure that the scaling with $H$ of the anisotropy as due to the large scale distribution of sources is different from that due to fluctuations. In particular the former contribution increases appreciably with increasing halo size, while the small-scale term is left more or less unaltered: as a result the small scale contribution, while comparable to $\delta_{A1}$ for values of $H$ that are smaller than the scale of the gradient in the source distribution, $R_{\odot}/\beta\sim 2.5$ kpc (see Eq.~\ref{eq:radial}), becomes close to negligible for $H=10$ kpc.

\begin{figure}[t]
\centering\leavevmode
\includegraphics[width=2.8in,angle=0]{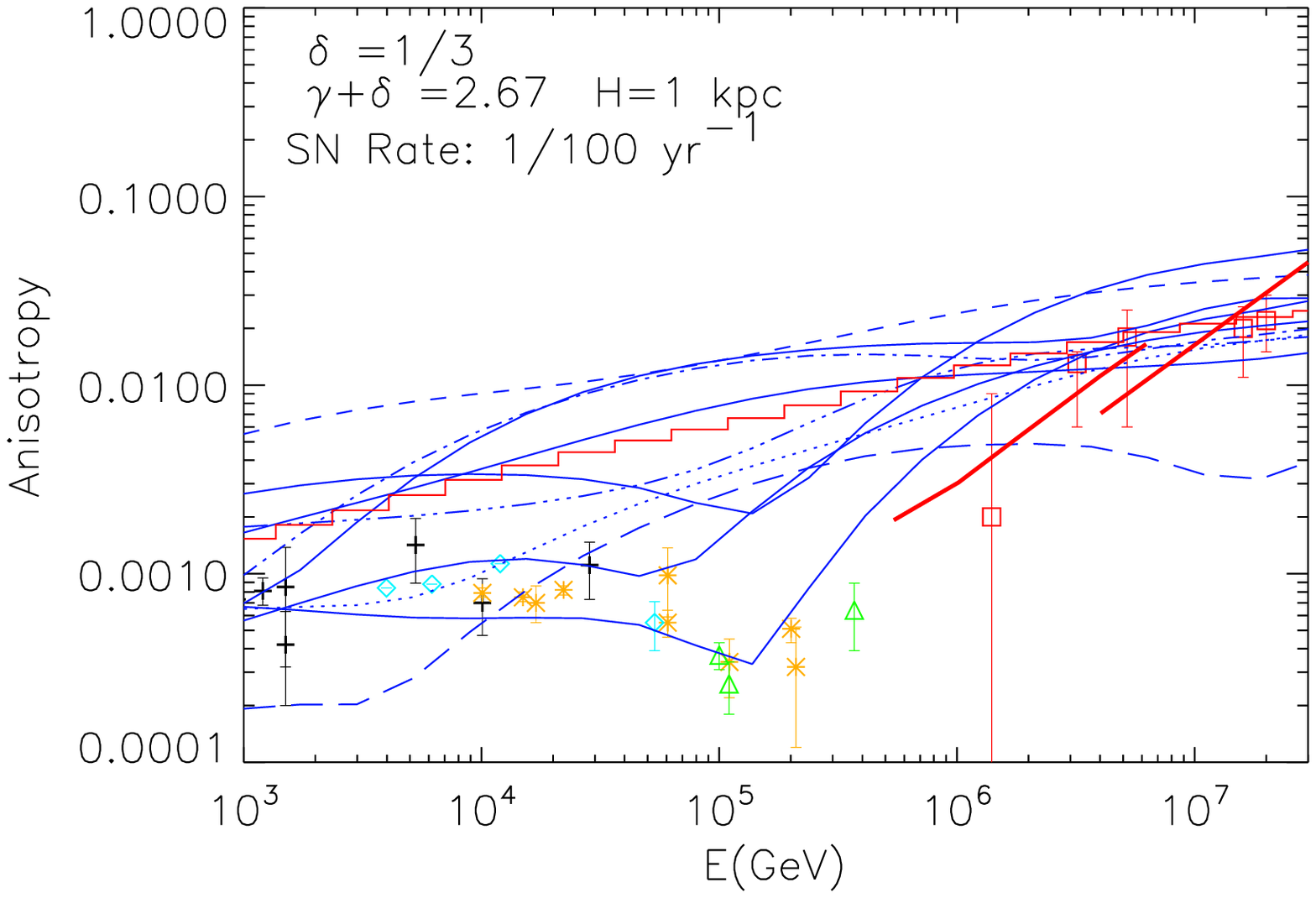}
\includegraphics[width=2.8in,angle=0]{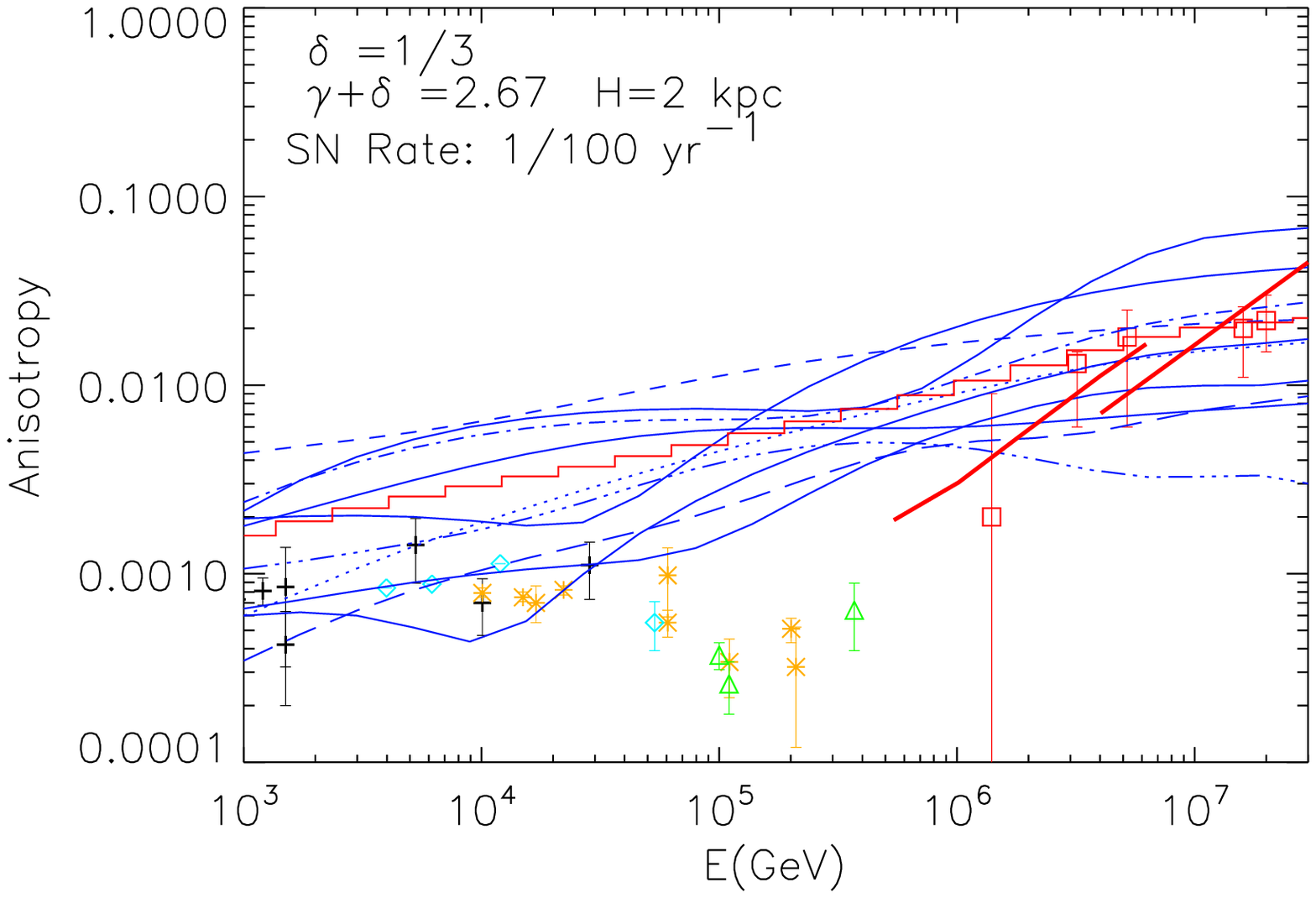}
\caption{Anisotropy amplitude for ten random realizations of sources in the cylindrical model for $H=1$ kpc (left panel) and $H=2$ kpc (right panel).}
\label{fig:varyH}
\end{figure}

In order to illustrate this effect within our model of stochastic source distribution we recalculate the amplitude of the anisotropy for $\delta=1/3$ and $\cR=1/100$ yr$^{-1}$ using a smaller halo size with $H=1$ and $H=2$ kpc. The results are illustrated in Fig.~\ref{fig:varyH}: both panels refer to the cylindrical model, with $H=1$ kpc on the left and $H=2$ kpc on the right. One can easily realize that increasing the size of the halo reduces the fluctuations. This can be understood because, as we stressed above, for large values of $H$ the anisotropy amplitude is dominated by the regular, large scale, distribution of the sources. In both cases, most realizations lead to an anisotropy that mostly grows with energy. In other words, for large values of $H$ it becomes harder to explain the anisotropy signal observed at Earth. In this sense, though not ruled out, large values of $H$ appear to be disfavored. Interestingly enough, in the case with $H=1$ kpc, there are a few realizations that lead to an anisotropy which is remarkably similar to the observed one.

\begin{figure}[t]
\centering\leavevmode
\includegraphics[width=2.8in,angle=0]{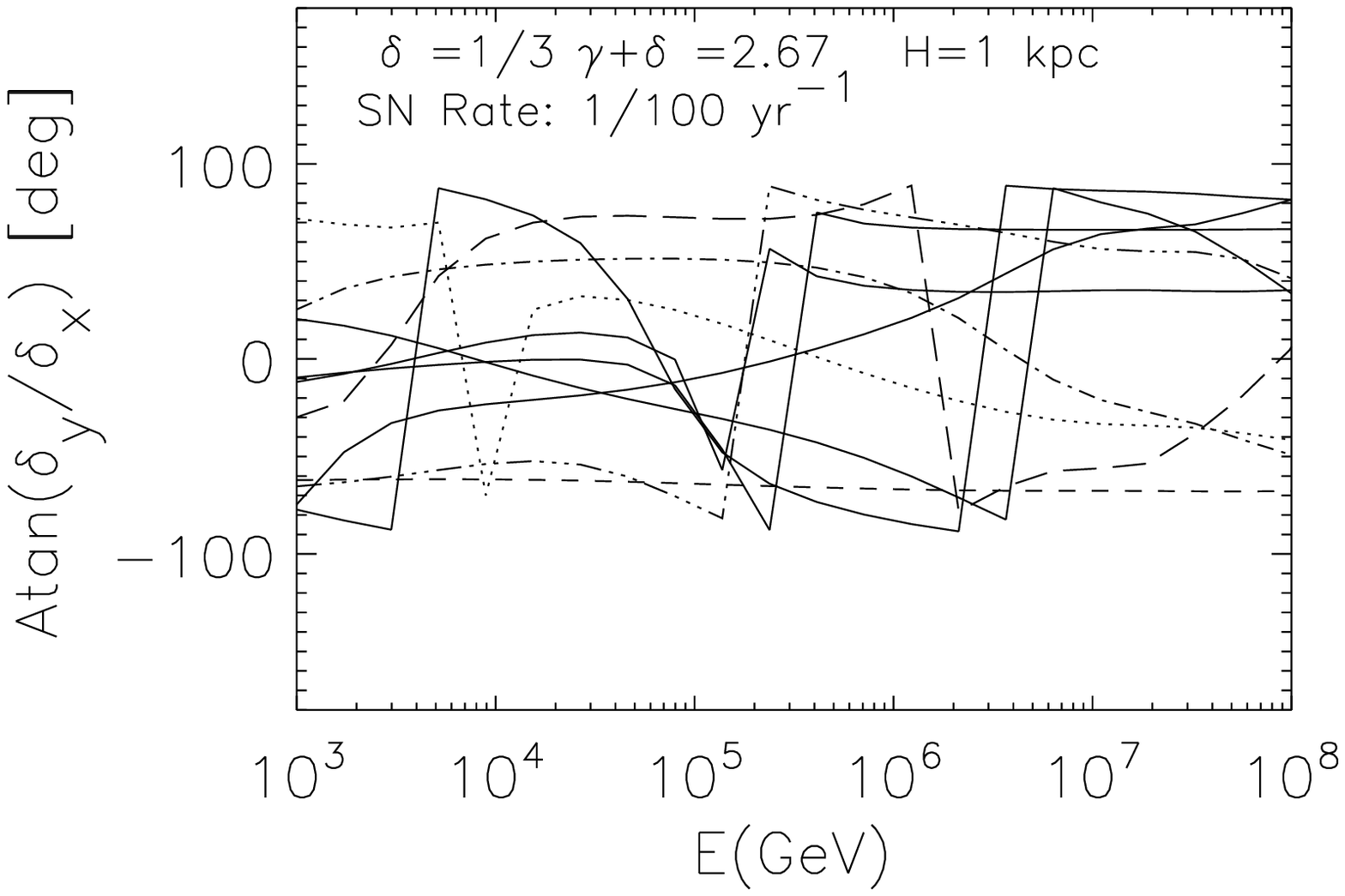}
\includegraphics[width=2.8in,angle=0]{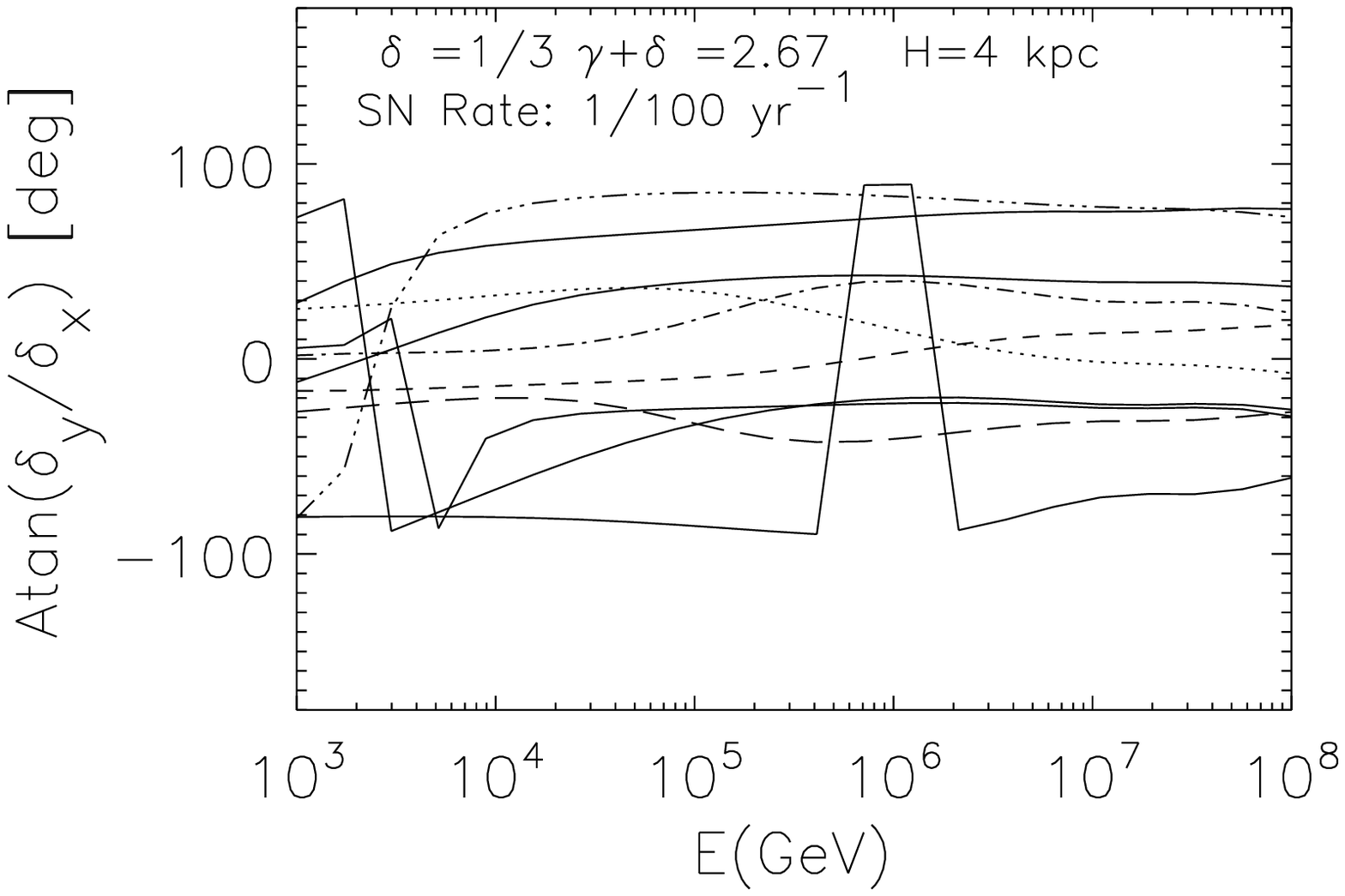}
\caption{Phase of the anisotropy in a cylindrical model with $H=1$ kpc (left panel) and $H=4$ kpc (right panel).}
\label{fig:varyHphase}
\end{figure}

The trend just described can be illustrated more clearly by using the phase of the anisotropy vector, as plotted in Fig.~\ref{fig:varyHphase} for a halo of size $H=1$ kpc (left) and $H=4$ kpc (right). For small values of $H$ the phase varies wildly reflecting the occasional dominance of a nearby recent source. Again, this behavior is reminiscent of that found by the EASTOP experiment \cite{Aglietta:2009p130}, as discussed above. For $H=4$ kpc the main contribution to anisotropy comes from the inhomogeneous source distribution in the Galactic disc, and the energy dependence of the phase of the anisotropy becomes much more regular, with an offset with respect to zero that reflects the presence of some nearby source.  

\section{Conclusions}\label{sec:conclusion}

We determined the amplitude of the anisotropy expected in the CR flux at Earth if SNRs are their main sources. We considered different scenarios in terms of the spatial distribution of Galactic SNRs, of the injection of CRs into the ISM from individual SNRs, and of the energy dependence of the diffusion coefficient $D(E)\propto E^{\delta}$. 

The amplitude of the anisotropy reflects the gradient in the CR density at Earth and includes both a contribution from the large scale distribution of sources in the Galaxy, and from the random nature (in space and time) of nearby recent SNRs. Whether the first or the second contribution dominates depends on the specific source distribution in space and on the choice of the halo size. For values of $H<4$ kpc the predicted anisotropy is the result of comparable contributions from the fluctuations induced in the CR flux from nearby sources and the inhomogeneous source distribution. This implies that the curve describing the anisotropy as a function of energy is characterized by bumps and dips, and is, in general, all but monotonic. In other words the naive and often quoted expectation that the anisotropy scales as $\propto D(E)$ is far from what one can measure for any specific realization of the source distribution. When the average of the anisotropy amplitude is computed over many source realizations, this turns out to be closer to the expected scaling, but the agreement is still not perfectly realized because of the presence of nuclei. These, after propagation, have spectra that are different, in general, from those of protons (see Paper I). As a result, the scaling of anisotropy with energy, as averaged over many source realizations, turns out to be somewhat slower than $\propto D(E)$. 

The bumps and dips mentioned above often resemble plateau regions in limited energy intervals, quite similar to what is actually observed at the Earth. Although it is not possible to explain the detailed structure of the anisotropy amplitude as reported in the figures above, since this depends upon the specific realization of nearby recent SNRs in which we live, we claim that the qualitative features of the observed data points can be understood as resulting from the stochastic nature of SNRs as sources of cosmic ray protons and nuclei. 

The most important conclusion of our calculations is that the anisotropy is a strong function of the diffusion coefficient. Assuming that the ratio $H/D(E)$ is normalized in such a way as to fit the observed ratio of B/C at $\sim 10~GeV/n$, we showed that the amplitude of the anisotropy for $\delta=0.6$ exceeds the observed amplitude by up to two orders of magnitude, depending on the realization of sources and on the energy. Larger values of $\delta$ make the problem even more severe. 

This is an important conclusion in that $\delta\geq 0.6$ would appear to be required by both the standard (test-particle) theory of particle acceleration at strong shocks (in which the slope of the injected spectrum is $\sim 2$) and by the non-linear version of the theory, in which at high energy it is usually found that the injection spectrum is even harder than $E^{-2}$ (see for instance \cite{Berezhko:2007p1010}). Only very recently \cite{Caprioli:2010p133,Ptuskin:2010p1025,Caprioli:2011p1915} it has been proposed that steeper spectra might result from phenomenological (and in general not very well justified in terms of basic physical principles) approaches to the problem of determining the velocity of the scattering centers. The fact that we find large values of $\delta$ to be incompatible with the measured amplitude of the anisotropy is an additional argument in favor of the existence of physical phenomena that steepen the injection spectrum, with all the implications that this finding involves.

All the conclusions relative to the choice of the diffusion coefficient remain valid also when the spiral structure of the Galaxy is taken into account. However, the spiral structure in the spatial distribution of SNRs affects the contribution to the amplitude of the anisotropy as due to the large scale gradient of CRs. The scale of the spiral arms is found to have an important effect on the anisotropy. If the distribution of the sources around the centroid of the arms is very narrow (case $w=2.8$ kpc above), the gradient induced by the spiral structure on the CR density dominates upon the fluctuations, hence the amplitude of the anisotropy grows with energy rather regularly, roughly as (but slightly more slowly than) $\propto D(E)$. For broader arms ($w=5$ kpc above), the role of fluctuations becomes dominant again, so that the amplitude shows the standard bumpy structure discussed earlier. 

In both cases of cylindrical and spiral distribution of SNRs in the Galaxy, the pattern of anisotropy is weakly affected by the size of the halo, $H$. However, when $H$ becomes much larger than the scale of gradients, namely $R_{\odot}/\beta$ in the cylindrical model, and $w$ in the spiral case, the contribution due to the CR gradient induced by the large scale source distribution dominates upon the fluctuating terms. Increasing the value of $H$ in general leads to a very regular trend of nearly monotonic growth of the anisotropy amplitude with energy, quite unlike what is observed. Such a behavior also reflects on the phase of the anisotropy which is very irregular for small $H$ and very regular for larger haloes. This argument plays against models with large halos, which however appear to be favored in calculations carried out with GALPROP. 

We conclude our discussion of the results of our calculations with a short comment on the importance of using a stochastic method like the one discussed here in order to have a reliable determination of the amplitude (and phase) of the anisotropy, rather than a propagation code such as GALPROP or DRAGON. These latter approaches can only pick the anisotropy (proportional to the CR gradient) induced by the large scale distribution of SNRs (or other sources) in the disc of the Galaxy, while missing completely the fluctuations, that however represent the dominant contribution to anisotropy for most values of the relevant parameters. At the same time it is fair to recall that both numerical propagation codes as well as the calculations presented here are not able to treat a major complication intrinsic to the problem, namely the fact that the propagation in the Galactic magnetic field might be more complex than modeled through a spatially constant, isotropic diffusion coefficient. For instance, it is very likely that the amplitude and phase of anisotropy are affected by anisotropic diffusion, namely preferential diffusion parallel to the magnetic field lines. On the other hand it is plausible that this effect is somewhat mitigated by the random walk of magnetic field lines, which leads to make diffusion close to isotropic. This also should reflect in a more erratic behaviour of the phase of the anisotropy vector, which seems pretty similar to what the data show. 

\section*{Acknowledgment}
We are grateful to Olivier Deligny, Paolo Desiati and Piera Ghia for providing help with the data points on the amplitude of anisotropy and to Damiano Caprioli, Giovanni Morlino, Rino Bandiera for continuous discussion on everything.

\bibliographystyle{JHEP}
\bibliography{crbib}

\end{document}